\documentclass[useAMS,usenatbib]{mnras}
\usepackage{graphicx}
\usepackage{longtable}

\title[The Veritas asteroid family]
      {Detection of the Yarkovsky effect for C-type asteroids in the Veritas
family}
      \author[V. Carruba, D. Vokrouhlick\'{y}, and D. Nesvorn\'{y}]{V. Carruba$^{1,3}$\thanks{E-mail: vcarruba@feg.unesp.br}, D. Vokrouhlick\'{y}$^{2}$, D. Nesvorn\'{y}$^{3}$\\
$^{1}$UNESP, Univ. Estadual Paulista, Grupo de din\^{a}mica Orbital e
  Planetologia, Guaratinguet\'{a}, SP, 12516-410, Brazil \\
$^{2}$Institute of Astronomy, Charles University, V Hole\v{s}ovi\v{c}k\'{a}ch 2, Prague 8, CZ-18000, Czech Republic\\
$^{3}$Department of Space Studies, Southwest Research Institute, Boulder, 
CO, 80302, USA\\
}

\begin{document}

\date{Accepted 2017 May 11.  Received 2017 May 11; in original form 2017 March 2.}

\pagerange{\pageref{firstpage}--\pageref{lastpage}} \pubyear{2017}

\maketitle

\label{firstpage}

\begin{abstract}
The age of a young asteroid family can be determined by tracking the orbits of 
family members backward in time and showing that they converge at some time in 
the past. Here we consider the Veritas family. We find that the membership  
of the Veritas family increased enormously since the last detailed analysis of 
the family. Using backward integration, we confirm the convergence of nodal 
longitudes $\Omega$, and, for the first time, also obtain a simultaneous 
convergence of pericenter longitudes $\varpi$. The Veritas family is found 
to be $8.23^{+0.37}_{-0.31}$~Myr old. To obtain a tight 
convergence of $\Omega$ and $\varpi$, as expected from low ejection 
speeds of fragments, the Yarkovsky effect needs to be included in the modeling 
of the past orbital histories of Veritas family members. Using this method, 
we compute the Yarkovsky semi-major axis drift rates, ${\rm d}a/{\rm d}t$, for
274 member asteroids. The distribution of ${\rm d}a/{\rm d}t$ values is
consistent with a population of C-type objects with low densities and low
thermal conductivities. The accuracy of individual ${\rm d}a/{\rm d}t$
measurements is limited by the effect of close encounters of member asteroids
to (1) Ceres and other massive asteroids, which cannot be evaluated with
confidence.    
\end{abstract}

%

\section{Introduction}
\label{sec: intro}

Asteroid families are the outcomes of disruptive collisions of main belt 
asteroids. After a family-forming event, the orbits of fragments are affected 
by gravitational and non-gravitational forces, such as planetary perturbations 
and the Yarkovsky effect \citep{Bottke_2002, Vokrouhlicky_2015}.  
The Yarkovsky effect acts to spread fragments in semi-major axis. It is 
therefore generally difficult, in case of old asteroid families, to 
distinguish between the original spread caused by ejection velocities 
and the subsequent evolution by the Yarkovsky effect.   

For young asteroid families (ages $<$20 Myr), on the other hand, the 
semi-major axis spread is not affected by the Yarkovsky effect. Their 
orbital structure thus allows us to make inferences about the original ejection
velocity field. This is why the young asteroid families are useful.
In addition, by integrating the orbits of young family members backward
in time and checking on the convergence of their longitudes of pericenter 
$\varpi$ and node $\Omega$, it is possible to determine the family's age 
\citep{Nesvorny_2003}.

The Veritas family (Family Identification Number, FIN, 609;
\citet{Nesvorny_2015}) was first studied by \citet{Milani_1994a}. They found
that the orbit of (490) 
Veritas diffuses chaotically in eccentricity due to a background mean motion 
resonance. For (490) Veritas to be classified as a member, the family must be
young ($<$50 Myr, \citet{Milani_1994a}). Subsequently, \citet{Nesvorny_2003} 
used the convergence of $\Omega$ of Veritas members to determine that the 
family is only 8.3 Myr old. This very young age was linked to a spike in
the terrestrial deposition of interplanetary dust particles at $8.2\pm0.1$ 
Myr ago \citep{Farley_2006}. 

Many asteroids have been discovered since 2003, and the population of Veritas 
members is now about ten times larger than it was back then.  Using techniques 
developed in \citet{Carruba_2016a}, here we investigate the interesting case 
of the Veritas family. Our goal is to: (i) revise the age estimate obtained in 
\citet{Nesvorny_2003}, (ii) show that the convergence constraint requires 
inclusion of the Yarkovsky effect in the backward integration, (iii) set 
constraints on values of the key parameters affecting the Yarkovsky force, 
such as the asteroids density and thermal conductivity, and (iv) study the 
effect close encounters with Ceres and other massive asteroids have had
on the past orbital histories of family members. 

The analysis of the Veritas family is complicated by the presence of the 
nearby 2:1 resonance with Jupiter, where the precession rate of the 
perihelion longitude, $g$, has a singularity. The precession rate $g$ is 
therefore fast in the region of the Veritas family, which prevented 
\citet{Nesvorny_2003} from demonstrating the convergence of $\varpi$. 
Here we were able to overcome this difficulty and obtain, for the first 
time, the simultaneous convergence of both $\Omega$ {\it and} $\varpi$. 
This increases our confidence that the present analysis correctly estimates 
the age of the Veritas family and constrains the principal parameters 
of the Yarkovsky effect. 

\section{Family identification and dynamical properties}
\label{sec: fam_ide}

As a first step of our analysis, we obtained the membership of the Veritas 
family from \citet{Nesvorny_2015}, where the family was defined using 
the Hierarchical Clustering Method (HCM, \citep{Bendjoya_2002}) and a 
cutoff of 30~m/s. 1294 members of the Veritas family were identified in
that work.  Following \citet{Carruba_2016} we identified objects in the 
local background of the Veritas family. For this, we used the database 
of synthetic proper elements available at the AstDyS site 
(http://hamilton.dm.unipi.it/astdys, \citet{Knezevic_2003}, accessed on
September 3, 2016).  Asteroids were considered to belong to the local
background if they had 
proper $e$ and $\sin i$ near the Veritas family, namely 
$0.035 < e < 0.095$ and $0.135 < \sin i< 0.185$ (these ranges correspond to 
four standard deviations of the observed distribution of the Veritas
family). 

\begin{figure*}
  \centering
  \begin{minipage}[c]{0.49\textwidth}
    \centering \includegraphics[width=3.1in]{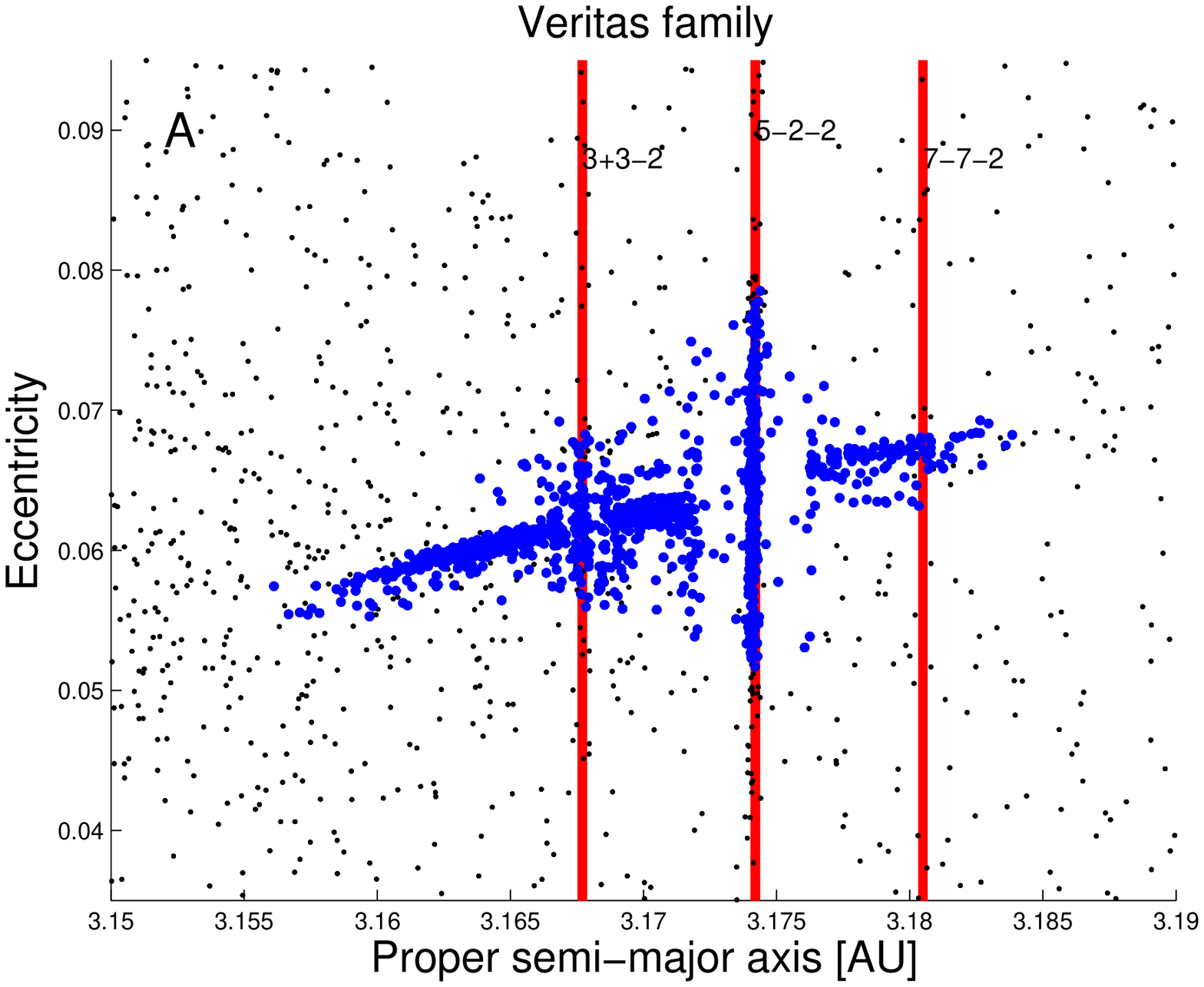}
  \end{minipage}%
  \begin{minipage}[c]{0.49\textwidth}
    \centering \includegraphics[width=3.1in]{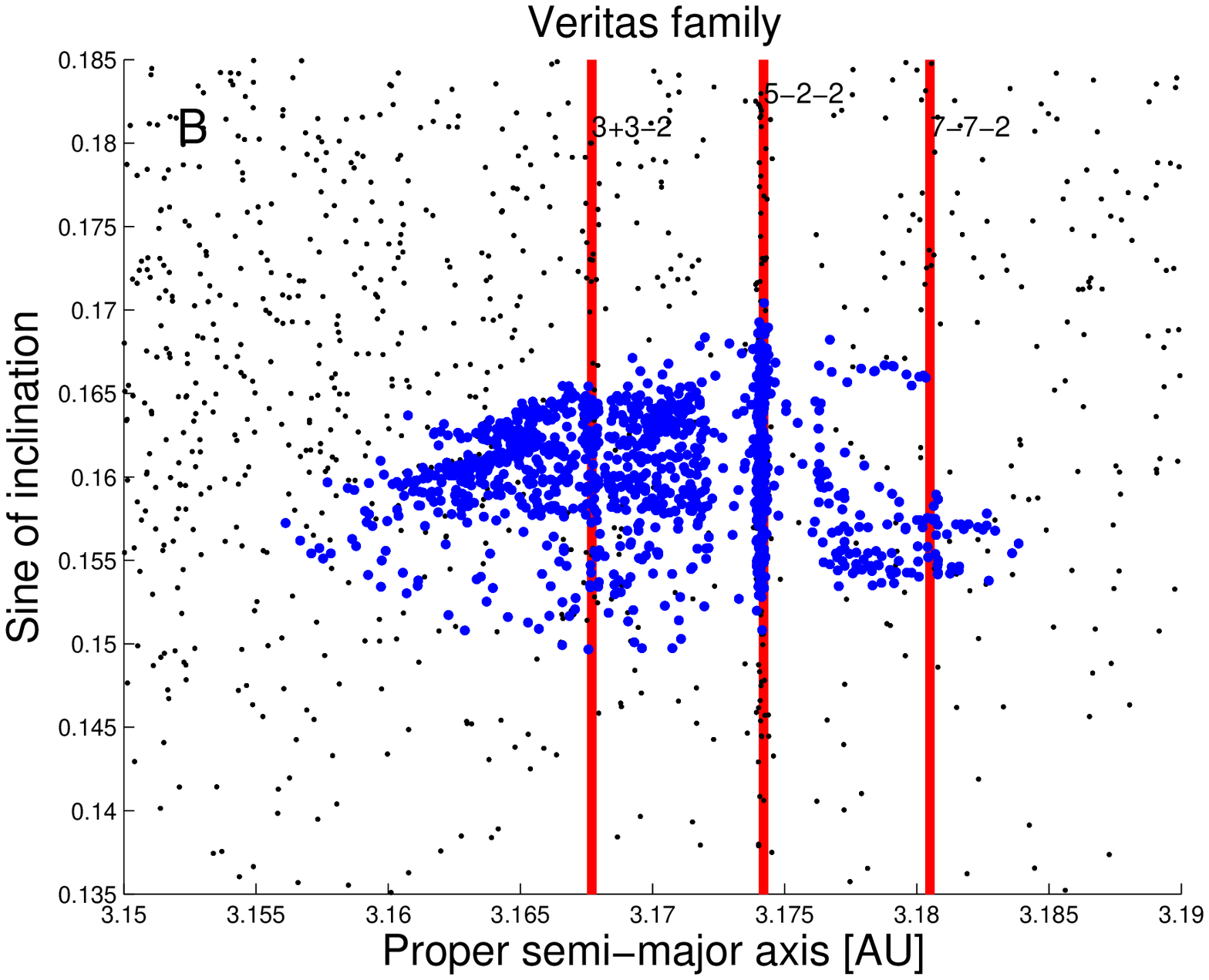}
  \end{minipage}
\caption{The $(a,e)$ (panel A) and $(a,\sin{i})$ (panel B) 
projections of orbits near the Veritas family. The vertical lines display
the locations of main mean-motion resonances in the region.  The blue 
symbols show the orbits of members of the Veritas family. The black dots 
display the background orbits.}
\label{fig: veritas_back}
\end{figure*}

The values of proper $a$ were chosen from the maximum
and minimum values of Veritas members plus or minus 0.02 au, the averaged
expected orbital mobility potentially caused by close encounters with 
massive asteroids over 4 Gyr \citep{Carruba_2013}.  Namely, this corresponds 
to an interval $3.15 < a < 3.19$ au.  Overall, we found 2166 background 
asteroids. No other important dynamical groups can be found in the 
local background of the Veritas family \citep{Carruba_2013b}.
After removing members of the Veritas family, the local background 
consists of 872 asteroids. The Veritas family and its background are shown 
in Fig.~\ref{fig: veritas_back}.

Chaotic dynamics in the region of the Veritas family was studied in detail
in \citet{Milani_1994a} and \citet{Tsiganis_2007}.  Interested readers can
find more information in those papers. Here we just consider the information 
about Veritas family members that can be obtained from their Lyapunov times.  
\citet{Tsiganis_2007} identified two main chaotic regions in the Veritas 
family: one, with Lyapunov times $<3 \times 10 ^4$~yr, associated with the 
three-body resonance 5-2-2 (or 5J:-2S:-2A in alternative notation) and its
multiplet structure, and another one, with $3 \times 10 ^4 < T_L < 10^5$ yr, 
caused by the interaction of asteroids with the 3+3-2 resonance 
(or 3J:3S:-2A). Another three-body resonance identified in the region 
was the 7-7-2 resonance, but that only affected a single asteroid 
((37005) 2000 TO37).  

Objects with Lyapunov times longer than $10^5$ yr and semi-major axes lower 
than that of the 3+3-2 resonance were classified as $R_1$ objects, while 
regular asteroids between the 3+3-2 and 5-2-2 resonances were classified 
as $R_2$ asteroids \citep{Tsiganis_2007}.  Only one regular Veritas family
member was known with semi-major axis larger than that of the 5-2-2 resonance.
A significant population of objects in this region is, however, currently
known (139 asteroids, Fig.~\ref{fig: veritas_back}). Extending the notation
from \citet{Tsiganis_2007}, we define these asteroids as being $R_3$ objects.

Concerning secular resonances in the region of the Veritas family,
an extensive study of the secular dynamics was performed
in \citet{Carruba_2013b, Carruba_2014}.  The two main secular
resonances near the Veritas family are the $g-2g_6+g_5+s-s_7$
(or, in terms of the linear secular resonances arguments, 
$2{\nu}_6-{\nu}_5+{\nu}_{17}$) and $g-g_6+2s-2s_6$ (or ${\nu}_6+2{\nu}_{16}$)
resonances. Only 10 outer main belt asteroids were found to librate
in these resonances. They do not thus play a significant role in the dynamical 
evolution of the Veritas family.

\begin{figure}
\centering
\centering \includegraphics [width=0.45\textwidth]{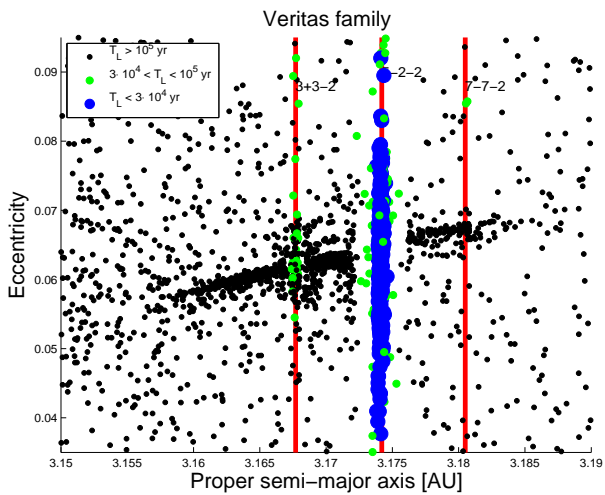}

\caption{The $(a,e)$ projection of orbits in the local background of
  the Veritas family.  Objects with Lyapunov times $T_L>10^5$
  yr are shown as black dots, objects with $3 \times 10 ^4 < T_L < 10^5$ yr
  are displayed as green full circles, and objects with 
$T_L < 3 \times 10^4$ yr are shown as blue full circles. The Lyapunov times 
were obtained from the AstDyS catalog (\citet{Knezevic_2003}). }
\label{fig: Lyap_ae}
\end{figure}

Fig.~\ref{fig: Lyap_ae} shows the $(a,e)$ projection of asteroids near the 
Veritas family.  The color code identifies the degree of chaoticity associated 
with a given orbit: regular orbits are shown as black dots, orbits with 
$3 \times 10^4 < T_L < 10^5$ yr are shown as green full circles, and orbits 
with $T_L < 3 \times 10 ^4$ yr are shown as blue full circles. With the
exception of the new population of regular objects beyond the 5-2-2 resonance,
our analysis essentially confirms that of \citet{Tsiganis_2007}.  

\section{Physical properties}
\label{sec: phys_prop}

A detailed analysis of physical properties of asteroids in the region of the 
Themis, Hygiea and Veritas families was reported in \citet{Carruba_2013b}.  
Here we briefly summarize the physical properties of asteroids near the 
Veritas family. There are 146 objects with photometric data in the
Sloan Digital Sky Survey-Moving Object Catalog data (SDSS-MOC4;
\citet{Ivezic_2001}) in this region, 89 of which (61\% of the total) are
members of the Veritas family.  In addition, 784 objects have geometric albedo
and absolute magnitude information 
available in the WISE and NEOWISE databases \citep{Masiero_2012}.

\begin{figure*}
  \centering
  \begin{minipage}[c]{0.49\textwidth}
    \centering \includegraphics[width=3.1in]{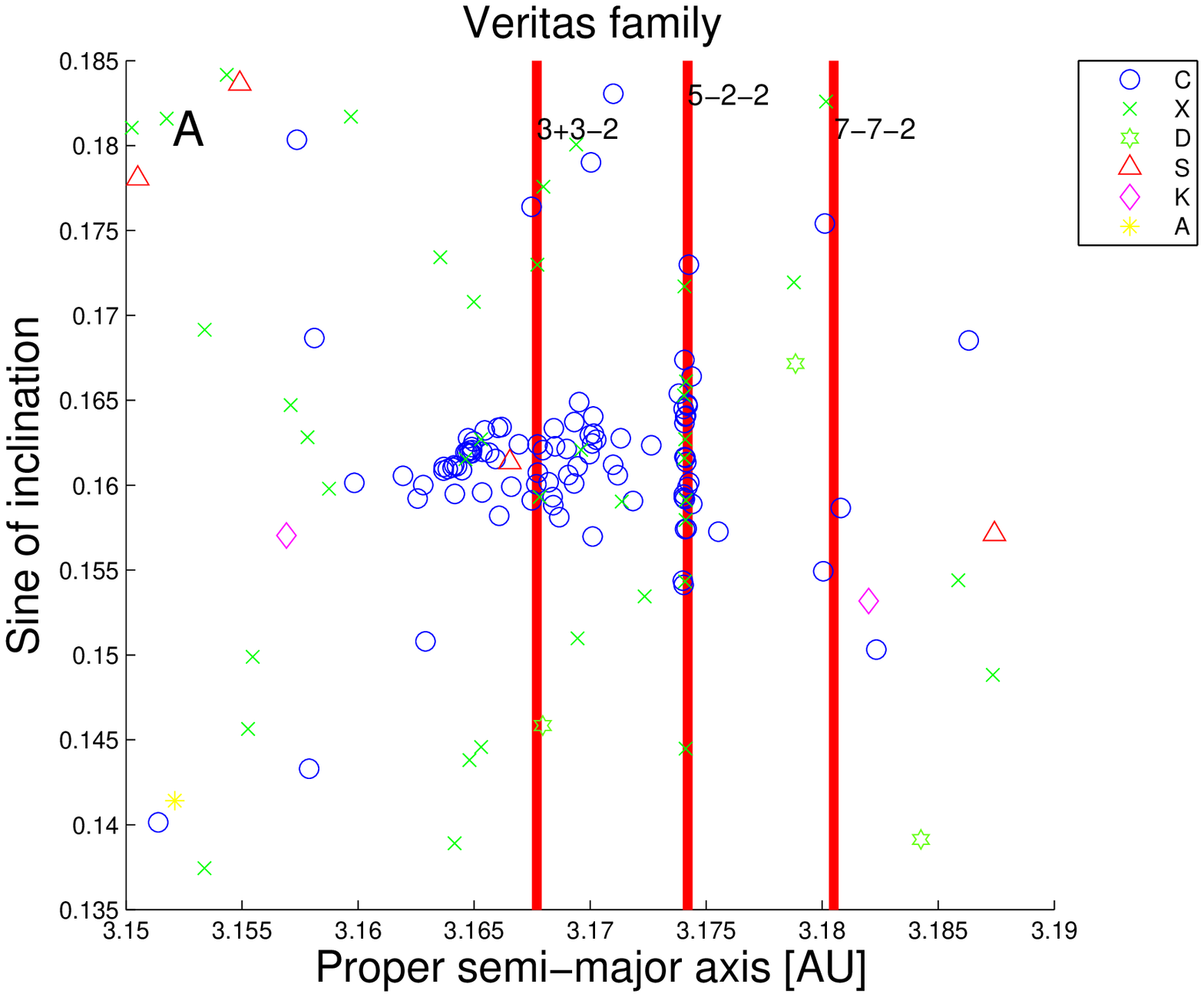}
  \end{minipage}%
  \begin{minipage}[c]{0.49\textwidth}
    \centering \includegraphics[width=3.1in]{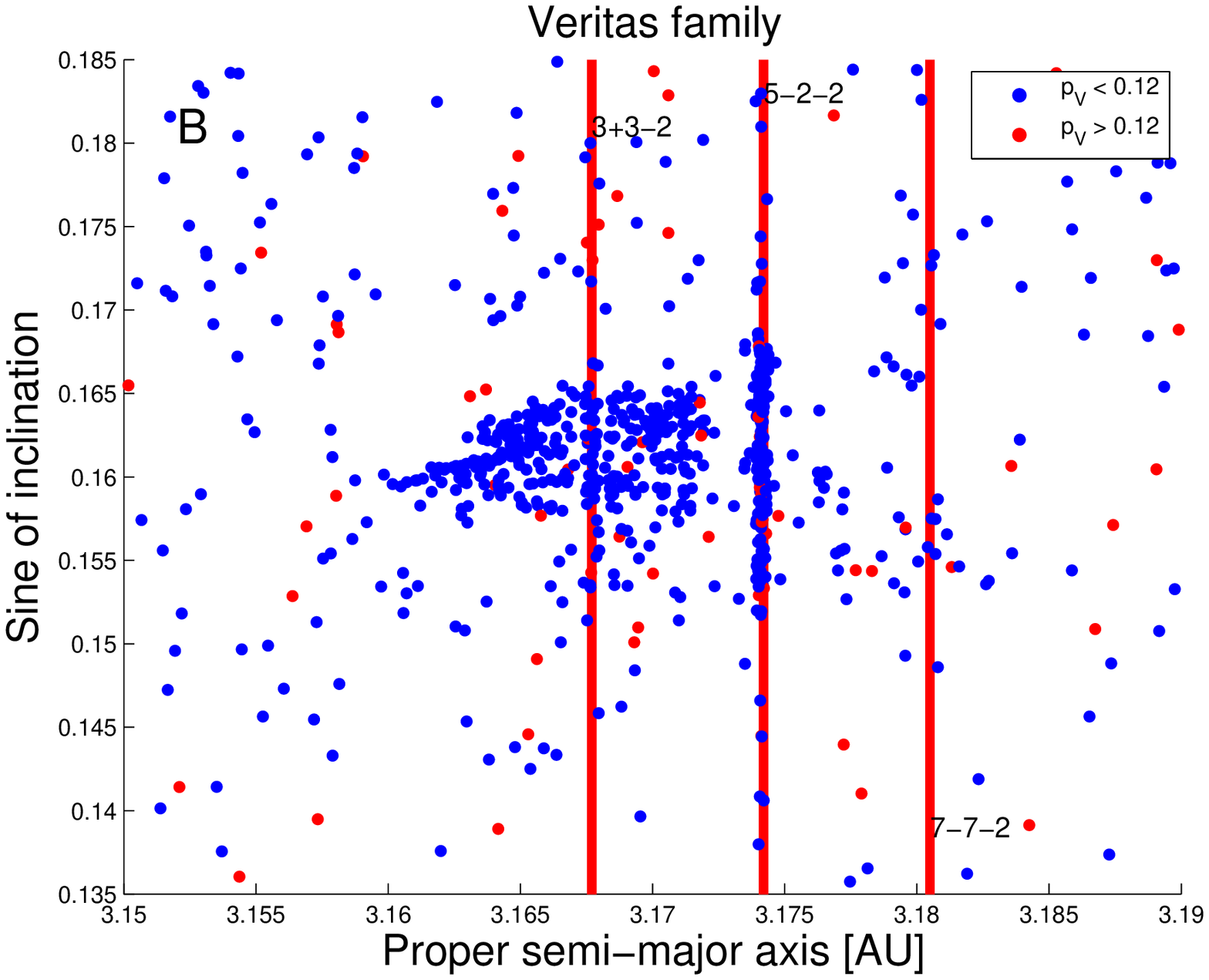}
  \end{minipage}
  \caption{An $(a,\sin{i})$ distribution of Veritas family asteroids,
    with taxonomic information (panel A) and WISE albedo data (panel B).
    Symbols used to identify asteroids with different spectral types
  and albedo values are identified in the inset.}
\label{fig: phys_prop}
\end{figure*}

Fig.~\ref{fig: phys_prop} shows the taxonomic classification of asteroids
obtained from SDSS-MOC4 with the method of \citet{DeMeo_2013} (panel A), 
and the WISE geometric albedo $p_V$ (panel B). In panel B, we separate dark 
asteroids (compatible with the C-complex taxonomy; $p_V < 0.12$) and
bright asteroids (S-complex taxonomy; $0.12 < p_V < 0.30$)
\citep{Masiero_2012}.  The Veritas family is obviously a C-type family, and
the C-complex objects 
also dominate the local background. In total, we found 97 Cs, 39 Xs and 3 Ds 
in the region, all belonging to the C complex.  There were only 7 S-complex 
asteroids, 4 of which are S-type, 2 K-type, and 1 A-type.  The proportion of 
C- and S-complex asteroids is consistent with the available geometric albedo 
data:  of the 784 objects with WISE albedo, 715 (91.2\% of the total) have 
$p_V <0.12$, and are compatible with a C-complex taxonomy.

Finally, we estimated the masses of asteroids in and near the Veritas family 
assuming objects to be spherical with bulk density equal to 1300 kg m$^{-3}$
(typical value of C-type objects).  For objects with available WISE albedo 
data, we used the WISE $p_V$ value to estimate their radius from the 
absolute magnitude (Eq.~1 in \citet{Carruba_2003}). For all other objects we 
used $p_V = 0.07$, which is the mean value of the Veritas family.

\begin{figure}
\centering
\centering \includegraphics [width=0.45\textwidth]{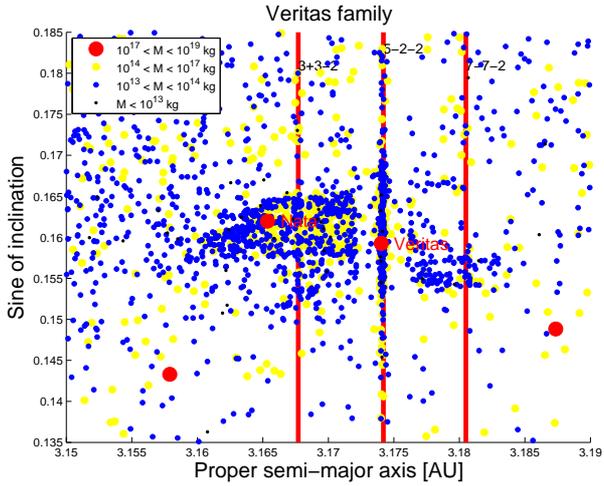}

\caption{An $(a,\sin{i})$ projection of asteroids near the orbital location of 
the Veritas family.  The color and size of the symbols reflect the estimated 
asteroid mass.}
\label{fig: masses}
\end{figure}

Fig.~\ref{fig: masses} shows our results.  Among the Veritas
members, only (490) Veritas and (1086) Nata have estimated masses larger than
10$^{17}$ kg and diameters $D>50$ km ($D=110$ km and $D=70$ km, respectively).
SPH simulation of the catastrophic disruption event that produced the
Veritas family \citep{Michel_2011} indicate the observed size distribution
of family members cannot be be well reproduced if both (490) Veritas and
(1086) Nata are true members of the family. Our analysis of the past convergence
described in the following sections shows that the convergence of 
(1086) Nata can be demonstrated, while that of (490) Veritas cannot (because 
of the chaotic orbit of (490) Veritas).  While it is not possible 
at this stage to positively decide whether (490) Veritas is a member or not of
its namesake family, for the purpose of our research we will use the orbit of
(1086) Nata as a reference for the method of convergence of secular angles
hereafter. Finally, since the combined volume of the Veritas
members, barring (490) Veritas itself, is about 40\% of the total volume of
the family, the family-forming event should be characterized as a
catastrophic disruption.

\section{Past convergence of the nodal longitudes}
\label{sec: long_conv}

Following the approach described in \citet{Nesvorny_2003, Nesvorny_2004,
Carruba_2016a} for the Veritas and Karin families, we checked for a 
past convergence of orbits of members of the Veritas family and
objects in the local background.  We first focus on demonstrating the 
convergence in $\Omega$, because the convergence of $\varpi$ is 
complicated by the proximity to the 2:1 resonance (both the $g$ 
frequency and its derivative, ${\partial g}/ {\partial a}$, are large)  
\citep{Nesvorny_2003}. 

To start with, to avoid strongly chaotic orbits, we selected 918 members 
of the Veritas dynamical family with $T_{\rm L}>3 \times 10 ^4$~yr. The 
chaotic orbits are mostly found in the identified three body resonances
and we cannot use them because their orbital histories cannot be computed 
deterministically. The selected orbits were integrated backward in time 
with $SWIFT\_MVSF$, which is a symplectic integrator programmed by
\citet{Levison_1994}.
It was modified by \citet{Broz_1999} to include online filtering of the 
osculating elements. All eight planets were included in the integration 
as massive perturbers. We used a time step of 1 day. The Yarkovsky effect
was not included in this initial integration. 

Using the approach described in \citet{Carruba_2016a}, we first
checked for the past convergence of $\Omega$. Asteroid (1086) Nata was used 
as a reference body, because its orbit has very long Lyapunov time
((490) Veritas cannot be used for this purpose because its orbit in the 
5-2-2 resonance is strongly chaotic). Specifically, we required that
$\Omega$ of individual orbits converge to within $\pm 60^{\circ}$ about
Nata's $\Omega$ in the time interval between 8.1 and 8.5 Myr ago, 
which encompasses the age of the Veritas family estimated in 
\citet{Nesvorny_2003}.  Out of the 918 considered bodies, 705 (76.8\% 
of the total) passed this test. 

We then turned our attention to objects in the Veritas family background.
First, as in \citet{Carruba_2016a}, we computed the terminal ejection 
velocities for the 705 bodies that passed the above criterion by 
inverting Gauss equations \citep{Murray_1999}:

\begin{small}
\begin{equation}
\frac{\delta a}{a} = \frac{2}{na(1-e^2)^{1/2}}[(1+e \cos{f}) \delta v_t +
e \sin{f}\delta v_{r}],
\label{eq: gauss_1}
\end{equation}

\begin{equation}
\delta e =\frac{(1-e^2)^{1/2}}{na}\left[\frac{e+2 \cos{f}+e \cos^2{f}}{1+e 
\cos{f}}\delta v_t+\sin{f} \delta v_r\right],
\label{eq: gauss_2}
\end{equation}

\begin{equation}
\delta i = \frac{(1-e^2)^{1/2}}{na} \frac{\cos{(\omega+f)}}{1+e \cos{f}} \delta v_W. 
\label{eq: gauss_3}
\end{equation}
\end{small}

\noindent 
where $\delta a = a-a_{ref}, \delta e = e-e_{ref}, \delta i= i-i_{ref}$, 
$a_{ref}, e_{ref}, i_{ref}$ define a reference orbit (we set $a_{ref}$ = 
3.170~au, $e_{ref}$ = 0.062 and $i_{ref}$ = 9.207$^{\circ}$) 
and $f$ and $\omega$ are the true anomaly and perihelion
argument of the disrupted body at the time of impact.  As in
\citet{Tsiganis_2007}, we used $f = 30^{\circ}$ and $\omega+f = 180^{\circ}$.

The highest terminal ejection velocities observed inferred from this exercise, 
excluding objects that obviously drifted away in the three-body resonances, was 
200 m/s.  We then integrated backward in time asteroids in the local background
of the Veritas family as defined in Sect.~\ref{sec: fam_ide} and eliminated 
objects that: (i) had Lyapunov times shorter than $3 \times 10 ^{4}$ yr, (ii)
had ejection velocities with respect to the reference orbit larger than 220 m/s
(i.e., 10\% larger than the maximum value determined above), and (iii) did not 
show the convergence of $\Omega$ to within $\pm 60^{\circ}$ around that of
(1086) Nata between 8.1 and 8.5 Myr ago. 

\begin{figure*}
  \centering
  \begin{minipage}[c]{0.49\textwidth}
    \centering \includegraphics[width=3.1in]{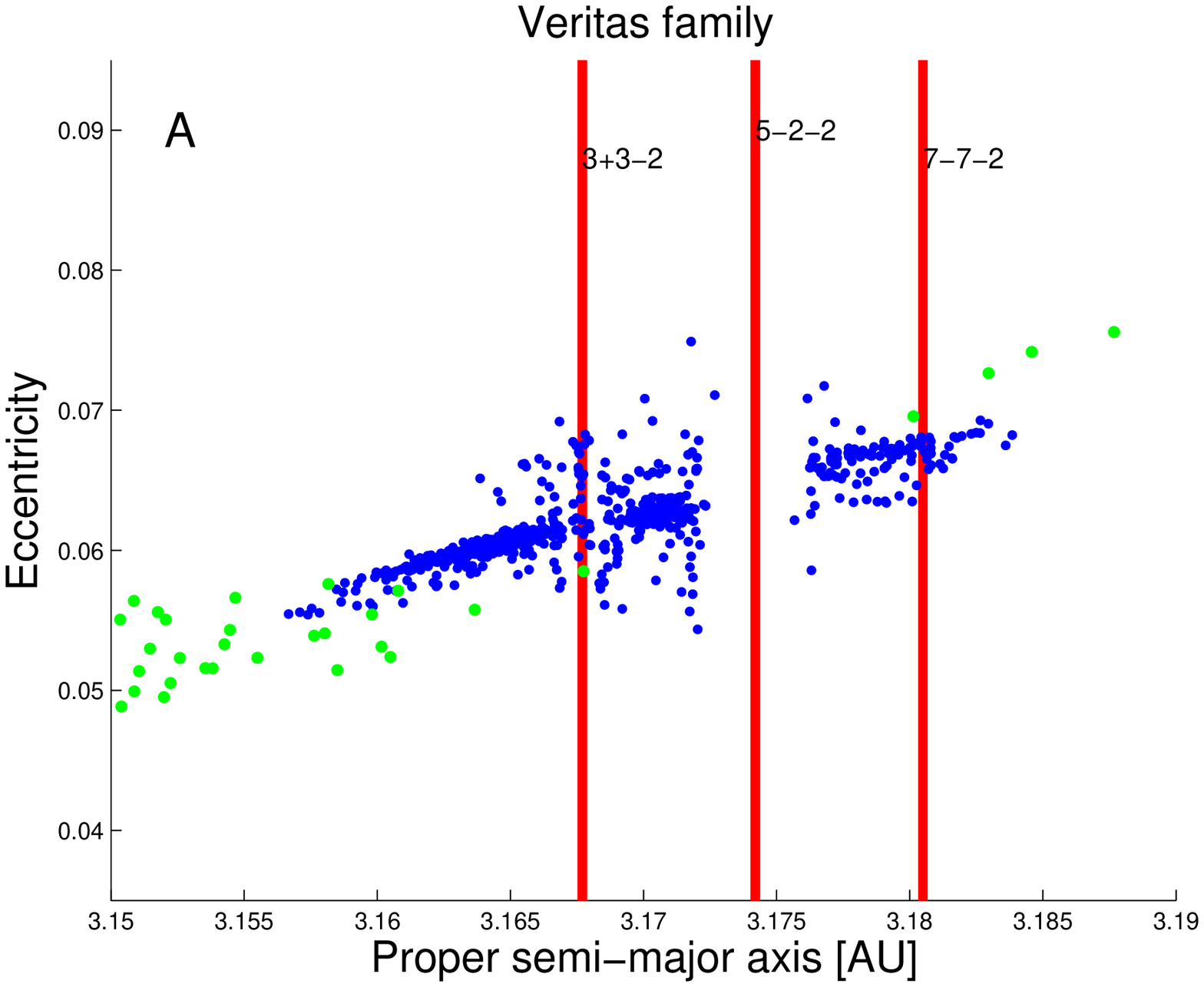}
  \end{minipage}%
  \begin{minipage}[c]{0.49\textwidth}
    \centering \includegraphics[width=3.1in]{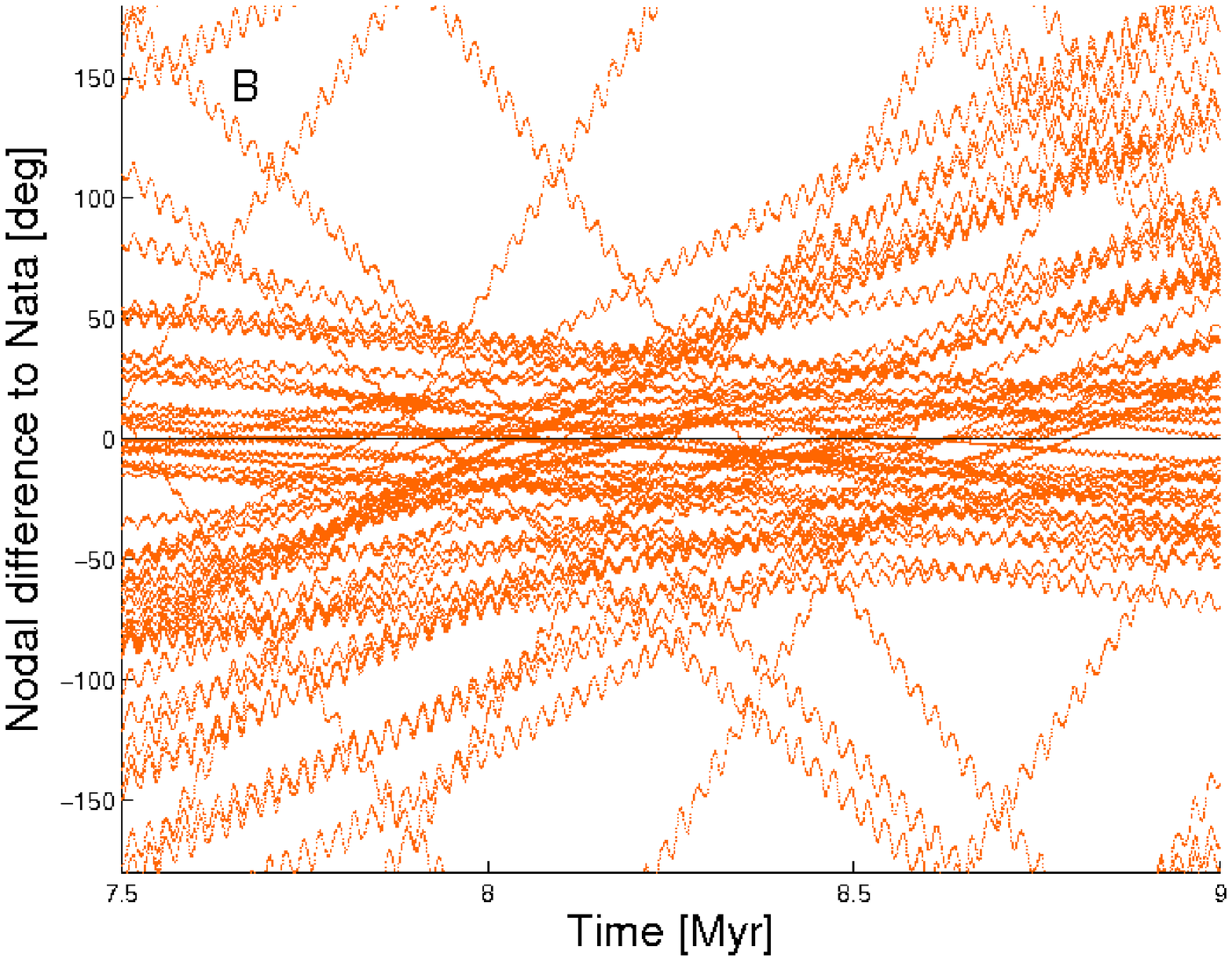}
  \end{minipage}
  \caption{Panel A: an $(a,e)$ projection of Veritas family members 
   (blue full circles) and background asteroids (green full circles) 
    that passed our convergence criterion. Panel B:
    convergence of the nodal longitudes at $\simeq$ 8.3 Myr of the first 
    50 members of the Veritas family (other members not shown for clarity). 
    The vertical dashed lines display 
    the approximate limits of the Veritas family age.}
\label{fig: Omega_conv}
\end{figure*}

Only 31 asteroids satisfied these requirements.  Since, however, most of these
objects were located at semi-major axis significantly smaller than those of the 
HCM members of the Veritas family (Fig.~\ref{fig: Omega_conv}, panel A), we
decided not to consider them for the following analysis.  After eliminating
taxonomical interlopers, we were left with 704 members of the Veritas
family.  Fig.~\ref{fig: Omega_conv}, panel A, shows the orbits of 704 members.
Panel B of that figure illustrates the convergence of nodal longitudes at
$\simeq$8.3 Myr ago.  

\section{Family Age and Detection of the Yarkovsky effect}
\label{sec: meas_drift}

The next step of our analysis is to obtain a preliminary estimate of the age of 
the Veritas family (as done in \citet{Carruba_2016a} for the Karin cluster),
and show the necessity to include the Yarkovsky effect in the backward 
integration in order to improve the convergence. We do this numerically.
The maximum Yarkovsky drift in $a$ for a 2~km C-type object, the 
smallest body in our Veritas sample, is roughly $2.0 \times 10^{-3}$~au over 
8.3 Myr (see \citet{Broz_2013} and Sect.~\ref{sec: drift_exp}). For
each of the 704 members of the Veritas family we therefore created 11 clones 
with the same initial orbits. Each clone was assigned a drift rate,
${\rm d}a/{\rm d}t$, from $-3.0 \times 10^{-10}$~au/yr to
$3.0 \times 10^{-10}$~au/yr, with a step of $0.6 \times 10^{-10}$~au/yr
between individual clones.

The limits of ${\rm d}a/{\rm d}t$ correspond to the maximum negative and
positive total drifts of $2.5 \times 10^{-3}$~au, i.e. about 25\% larger than
the maximum expected change in $a$ for the smallest fragment over the
estimated age of the family.  
All 7744 clones were then integrated backward in time over 10 Myr with 
$SWIFT\_RMVS3\_DA$, a symplectic integrator based on $SWIFT\_RMVS3$ code 
\citep{Levison_1994} that was modified by \citet{Nesvorny_2004} to include a 
constant drift in the semi-major axis.

\begin{figure}
\centering
\centering \includegraphics [width=0.45\textwidth]{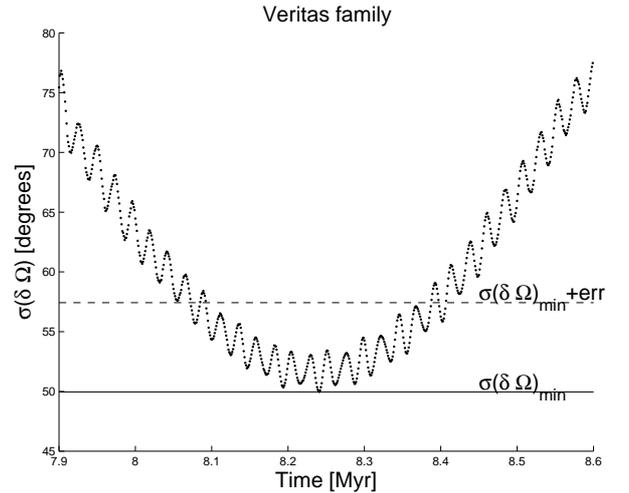}
\caption{The evolution of the standard deviation of $\Delta \Omega$ as a
  function of time.  The horizontal line identifies the minimum
  value of $\Delta \Omega$, while the dashed line shows the minimum value
  plus its error, assumed equal to one standard deviation of the
  $\Delta \Omega$ values over the length of our integration.}
\label{fig: da_com_veritas}
\end{figure}

The integration was used to refine the age estimate of the Veritas family.
Here we only used the past convergence of $\Omega$. For each time output 
of the integration between 7.9 to 8.6 Myr ago, we computed the standard 
deviation of $\Delta {\Omega}$ values (deltas computed with respect to (1086) 
Nata) for 704 clones of our simulation with {\it zero} Yarkovsky drift. We then 
searched for the minimum of the standard deviation of $\Delta {\Omega}$.
Figure~\ref{fig: da_com_veritas} displays the time evolution of
$\sigma(\Delta {\Omega})$ 
as a function of time for the simulated asteroids.  Based on this, the age of 
the Veritas family was found to be $8.24\pm0.17$~Myr.  

Adopting this age, we identified the value of ${\rm d}a$ for each clone 
that minimizes its $\Delta {\Omega}$. We found that the convergence in
$\Delta {\Omega}$ is still not perfect, partly because of the rough resolution
of ${\rm d}a/{\rm d}t$ with only 11 clones and partly because many near 
resonant orbits, which were not filtered out with our Lyapunov time cut, 
displayed significant chaos. To avoid these problems, we applied a narrower
selection of 274 objects, which: (i) have semi-major axes less than
3.166~au (to avoid possible interactions with the 3+3-2 resonance), and
(ii) have Lyapunov times greater than $2\times10^{5}$~yr, to avoid chaotic
orbits. Since one of our goals with this numerical experiment is 
to verify the possible past convergence of $\varpi$, we believe
that our approach based on selecting the most regular objects in the $R_1$
region, including (1086) Nata, is justified.  For each value of
${\rm d}a/{\rm d}t$ 
obtained from the previous simulations, we created 31 additional clones of the 
same particle with ${\rm d}a/{\rm d}t$ values covering plus or minus
the step value of $0.6 \times 10^{-10}$~au/yr used in the previous integration.
Overall, we integrated 8494 orbits. 

\begin{figure*}
 \centering
 \begin{minipage}[c]{0.49\textwidth}
    \centering \includegraphics[width=3.1in]{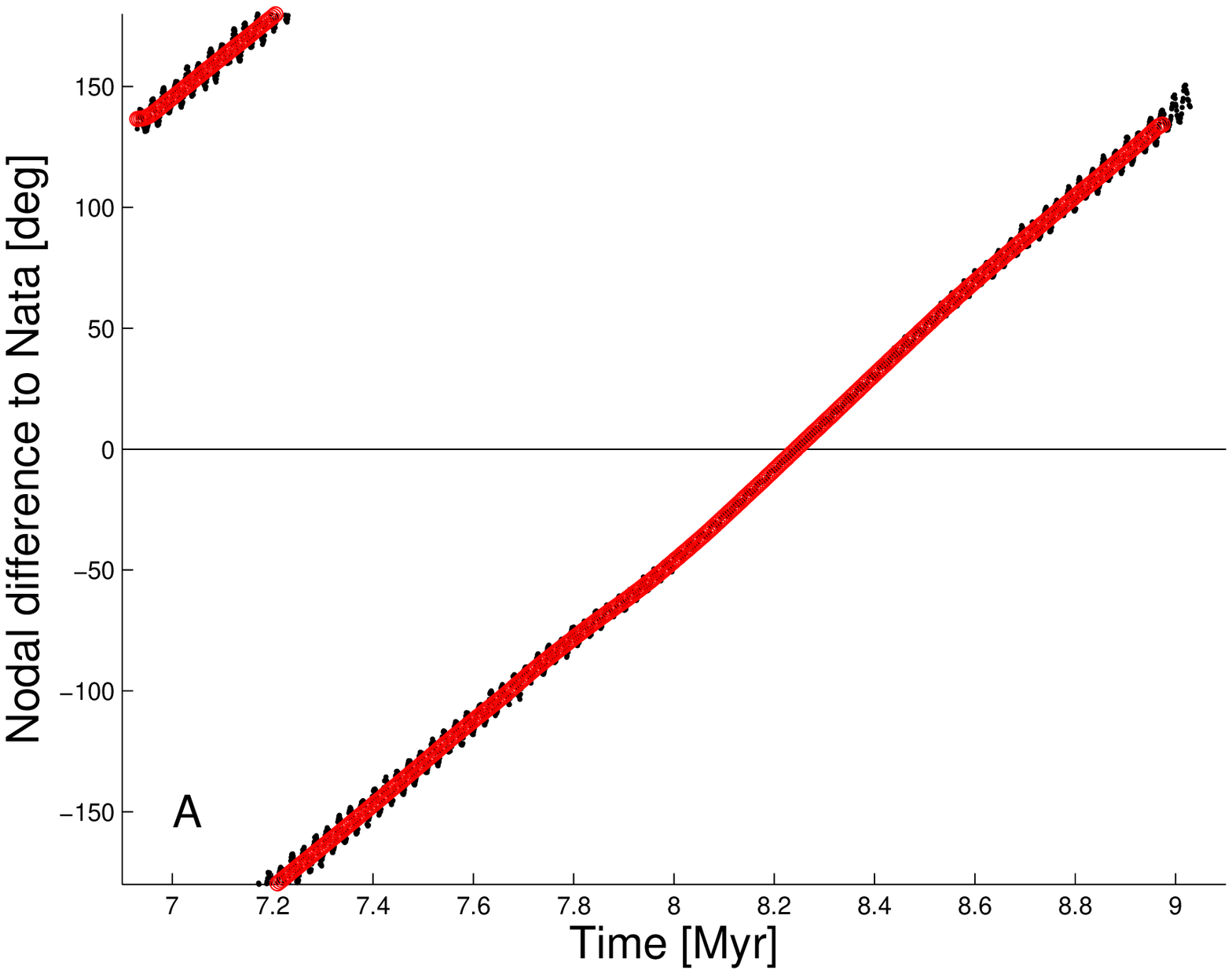}
  \end{minipage}%
  \begin{minipage}[c]{0.49\textwidth}
    \centering \includegraphics[width=3.1in]{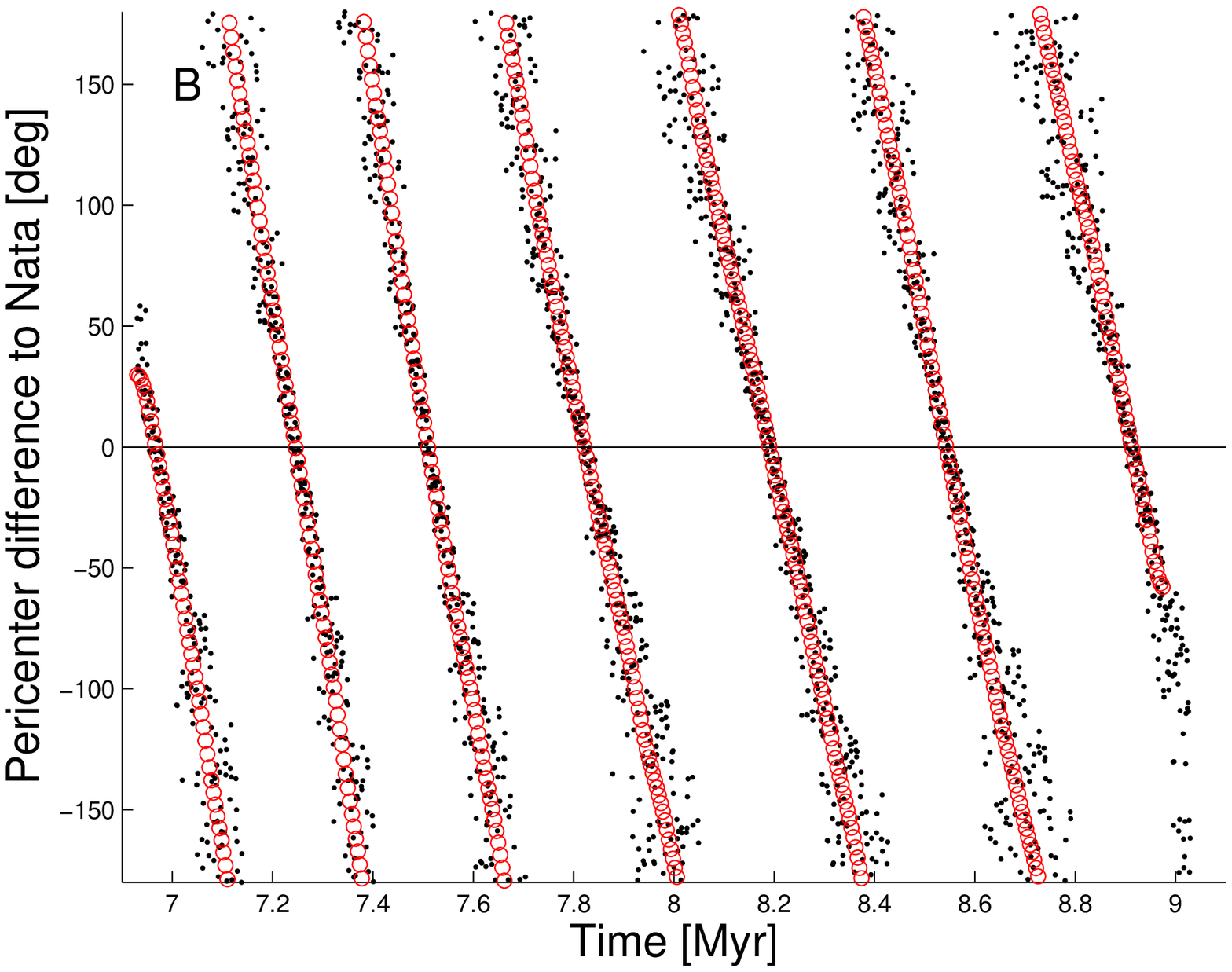}
  \end{minipage}

\caption{Osculating (black dots) and filtered (red circles) values
  of $\Delta \Omega = {\Omega}_{ast}-{\Omega}_{Nata}$ (panel A), and
  of $\Delta {\varpi} = {\varpi}_{ast}-{\varpi}_{Nata}$ (panel B), for one
  of the 8494 integrated particles in the second run.  All frequency terms
  with a period smaller than $10^{5}$~yr were removed in the filtered elements.}
\label{fig: filtered_el}
\end{figure*}

To better identify the orbits whose $\varpi$ angles 
converge to that of (1086) Nata, we filtered
$\Delta {\Omega}_{i}={\Omega}_{i}-{\Omega}_{Nata}$ and 
  $\Delta {\varpi}_{i}={\varpi}_{i}-{\varpi}_{Nata}$, where the subfix $i$
  indicates the i-th asteroid, with a low-pass digital
Fourier filter (see \citet{Carruba_2010} for a description of the filtering
method).  This removed all frequency terms with periods shorter than
$10^{5}$~yr. Fig.~\ref{fig: filtered_el} illustrates this procedure in
an example. 

We then analyzed the time behavior of the digitally filtered
  $\Delta {\Omega}$ and
$\Delta {\varpi}$ angles.  We first obtained a refined age estimate of the
Veritas family using two approaches: for each of the 31 clones of a given 
asteroid, we selected the one with the minimum values of
  $\Delta{\Omega}$ and $\Delta{\varpi}$ at each time step.  We then computed
${\chi}^{2}$-like variables using the relationships:

\begin{equation}
{\chi}^{2}_{1} = \sum_{i=1}^{N_{ast}}(\Delta {\Omega}_{i})^2,
\label{eq: chi2_age_nodes}
\end{equation}

\begin{equation}
{\chi}^{2}_{2} = \sum_{i=1}^{N_{ast}}[(\Delta {\varpi}_{i})^2+(\Delta {\Omega}_{i})^2],
\label{eq: chi2_age}
\end{equation}

\begin{figure*}
\centering
  \begin{minipage}[c]{0.49\textwidth}
    \centering \includegraphics[width=3.1in]{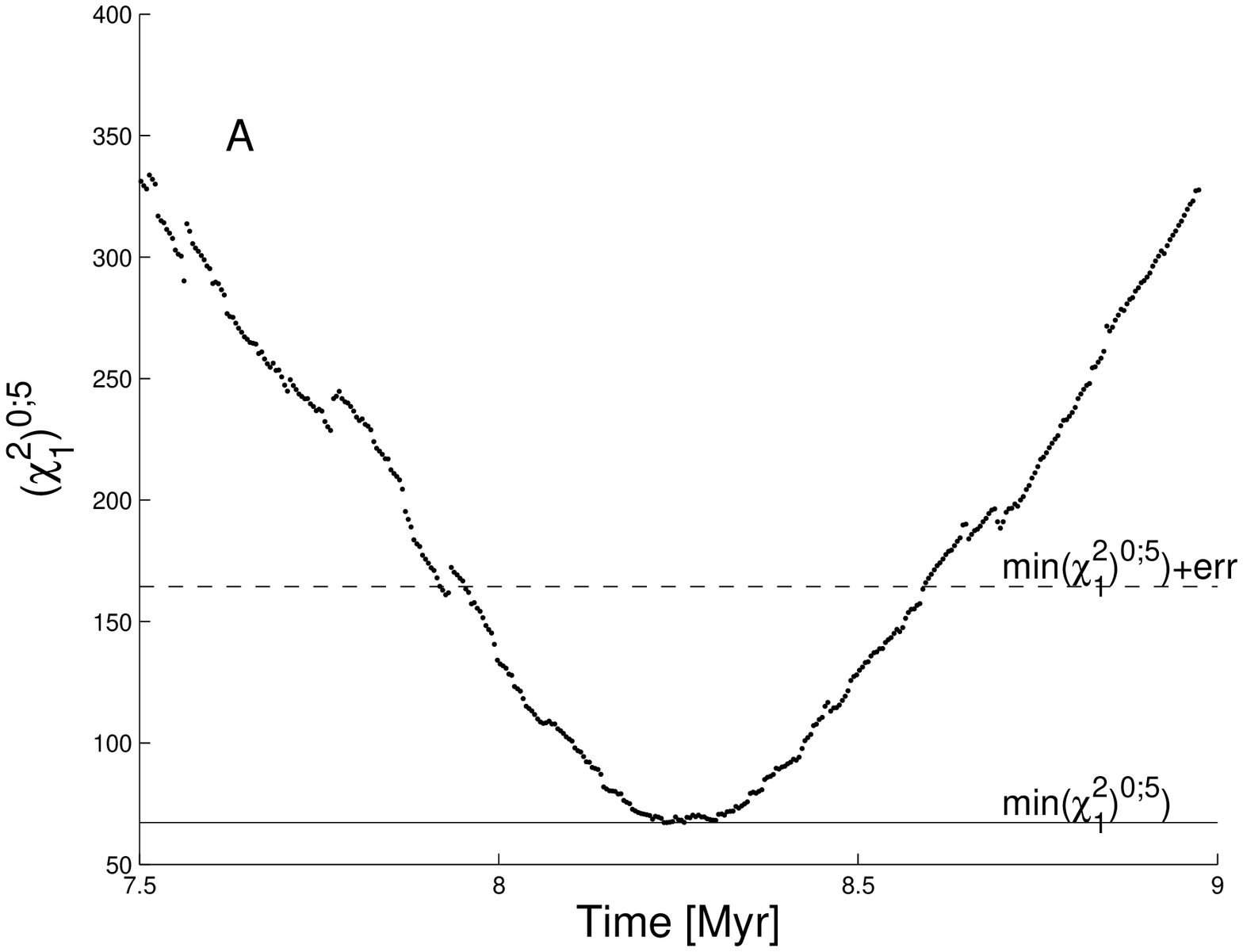}
  \end{minipage}%
  \begin{minipage}[c]{0.49\textwidth}
    \centering \includegraphics[width=3.1in]{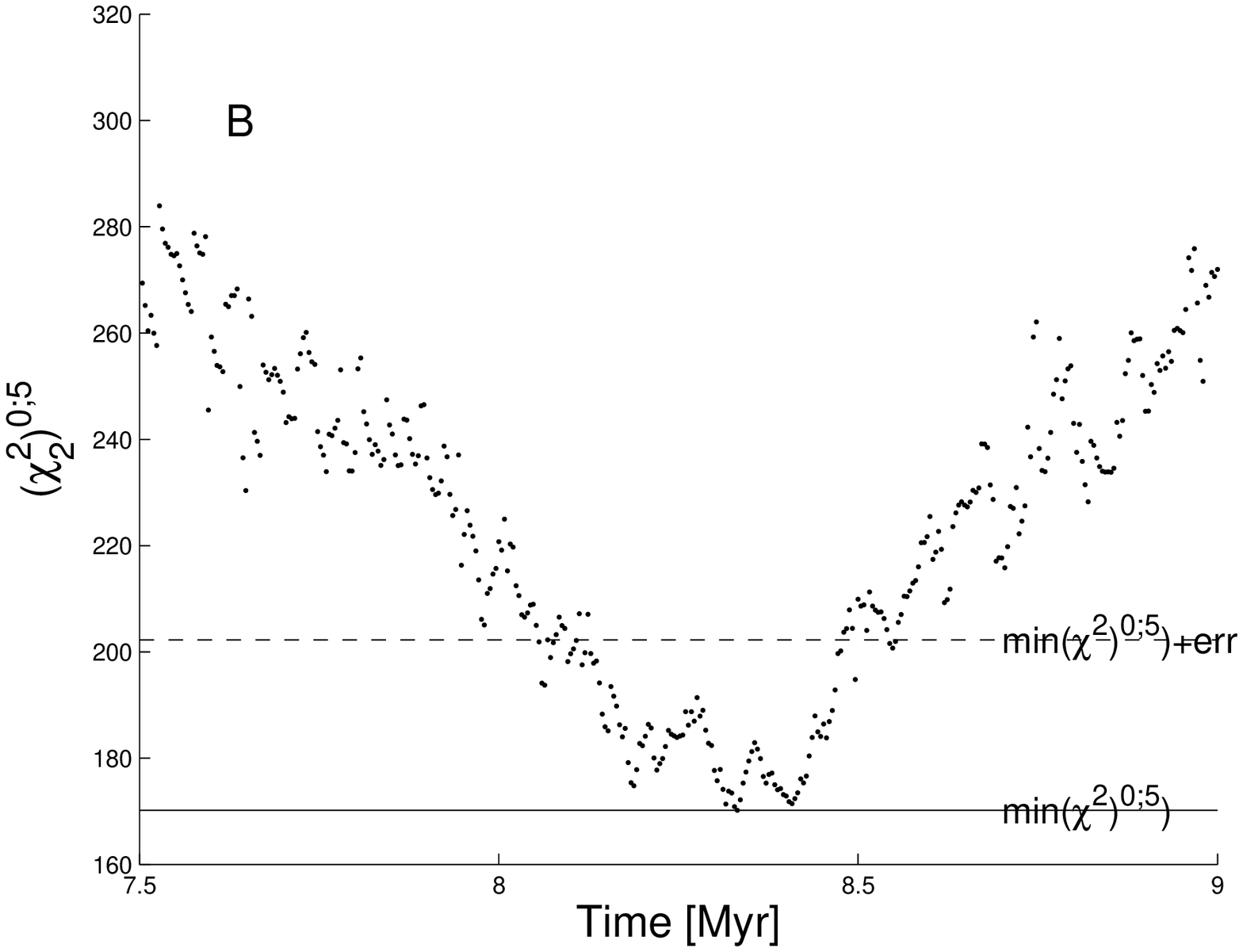}
  \end{minipage}
\caption{Time evolution of the square root of ${\chi}^{2}$, as defined
in Eqs.~(\ref{eq: chi2_age_nodes}) and (\ref{eq: chi2_age}).  The horizontal 
black line displays the minimum value of these quantities, while the dashed 
line show the minimum plus the error, defined as square root of the standard 
deviation of ${\chi}^{2}$ over the considered time interval.}
\label{fig: chi2_age}
\end{figure*}

\noindent
where $N_{ast} = 274$ is the number of asteroids in our sample. The 
first relationship will minimize the dispersion in $\Delta {\Omega}$, 
while the second will minimize the dispersion in both $\Delta {\Omega}$ and
$\Delta {\varpi}$.  Since the convergence in $\Delta {\Omega}$ is more robust, 
the first method provides a better estimate for the age.  We will
consistently used results from this first method hereafter. The second 
method, however, shows that convergence in $\Delta {\varpi}$ is actually
possible, at least in the numerical model here considered. 

\begin{figure*}
 \centering
 \begin{minipage}[c]{0.49\textwidth}
    \centering \includegraphics[width=3.1in]{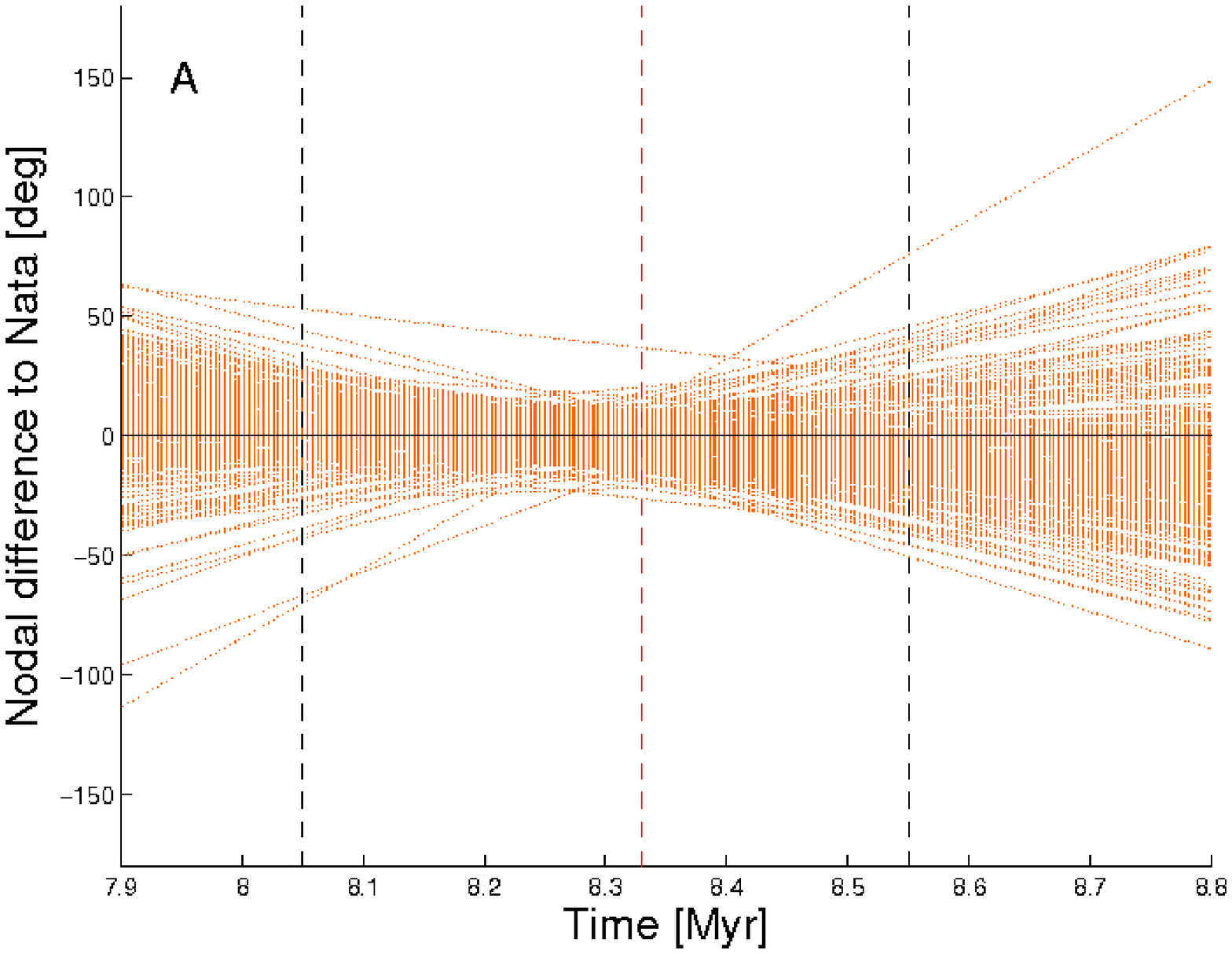}
  \end{minipage}%
  \begin{minipage}[c]{0.49\textwidth}
    \centering \includegraphics[width=3.1in]{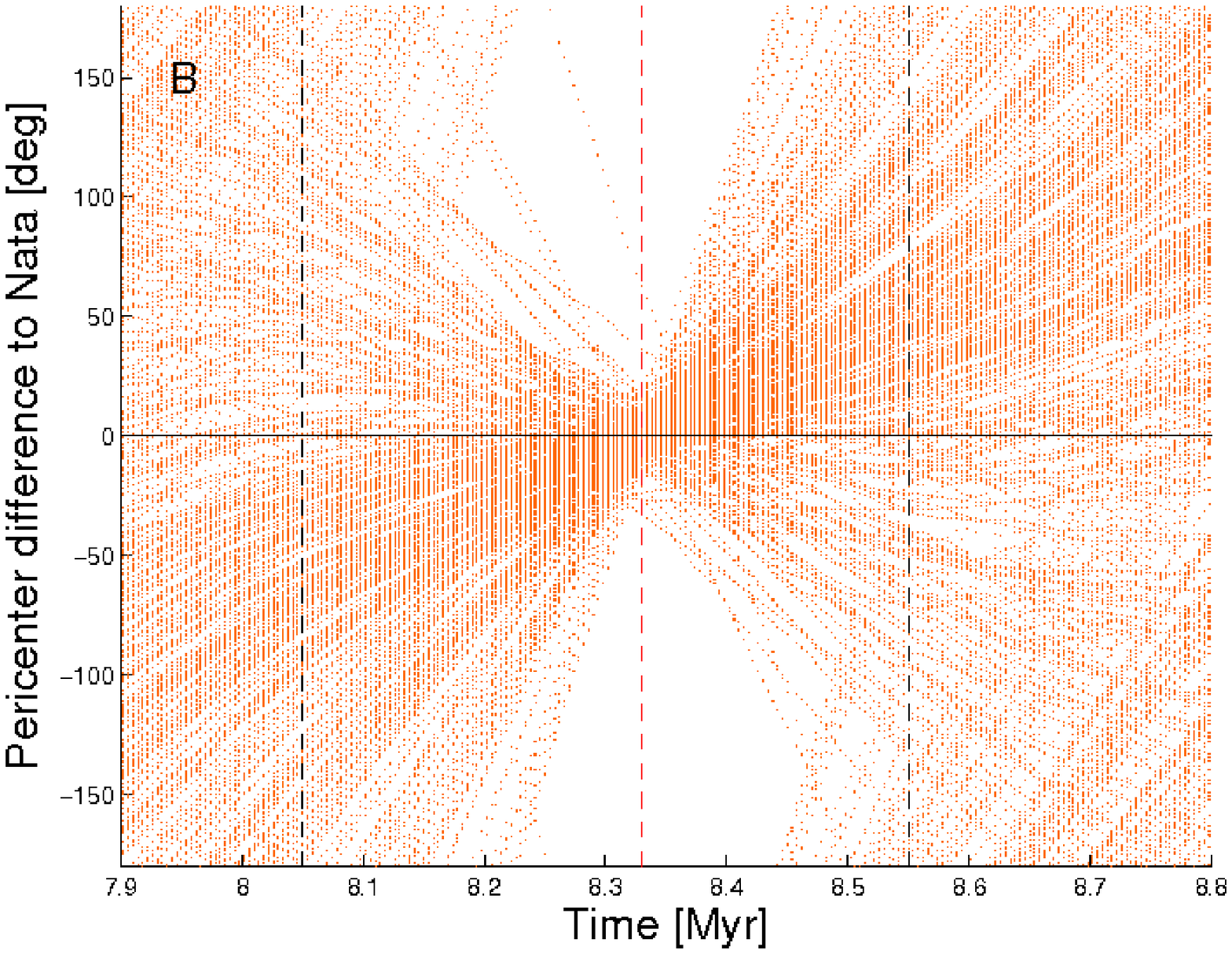}
  \end{minipage}

\caption{Convergence of filtered $\Delta {\Omega}$ (panel A) and
$\Delta {\varpi}$ (panel B)  for the 274 particles with the lowest values
of these angles at the nominal family age.  Vertical red line displays 
the nominal Veritas family age, while the vertical black dashed lines 
show the possible range of family ages obtained with the ${\chi}^2$-like 
approach of Eq.~(\ref{eq: chi2_age}).}
\label{fig: conv_filt_angles}
\end{figure*}

\begin{figure}
\centering
\centering \includegraphics [width=0.45\textwidth]{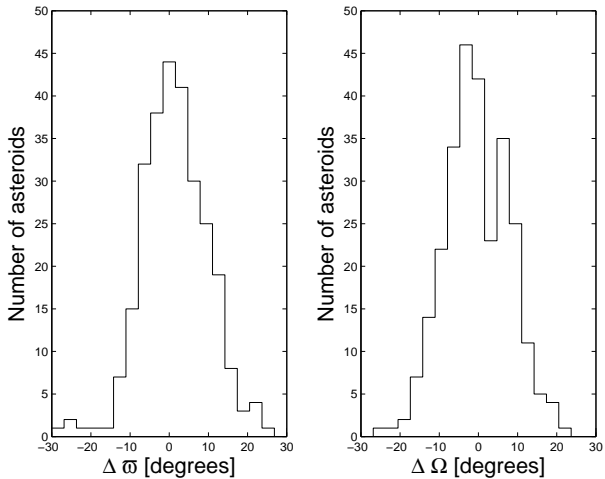}
\caption{Histograms of the distributions of $\Delta {\varpi}$ and
  $\Delta{\Omega}$ obtained at the family nominal age of 8.33 Myr, computed
  with our second approach.}
\label{fig: dom_dvarpi}
\end{figure}

Fig.~\ref{fig: chi2_age} displays the time
evolution of the square root of these two quantities (see panel A and B).  We
define the nominal error of $\sqrt{{\chi}^{2}}$ 
as the standard deviation of these quantities over the considered time
interval.  Based on this analysis, the Veritas family is
$8.23^{+0.37}_{-0.31}$~Myr old.   It is important to point out that, for 
the first time, we were also able to obtain the convergence of $\varpi$.  
At the nominal family age of $8.33^{+0.22}_{-0.28}$ Myr obtained with the
second method, we identified a set of 274 clones whose angles have minimum
values of $\Delta {\varpi}$ and $\Delta {\Omega}$.  The result is shown in 
Fig.~\ref{fig: conv_filt_angles}:  the age solutions of the Veritas
family with both methods are statistically identical.  Apart from internal
consistency, it also  justifies the convergence of the longitudes
of pericenter. Overall, the convergence of the angles is remarkable: the
standard deviations of the distribution in $\Delta {\Omega}$  and
$\Delta {\varpi}$ at the family nominal age are $8.4^{\circ}$ and $8.7^{\circ}$,
respectively.  Fig.~\ref{fig: dom_dvarpi} shows histograms of the two
distributions. The results obtained with the first method were similar: the
standard deviations in this case are equal to $7.7^{\circ}$ and $8.9^{\circ}$,
respectively.

\begin{figure}
  \centering
  \centering \includegraphics [width=0.45\textwidth]{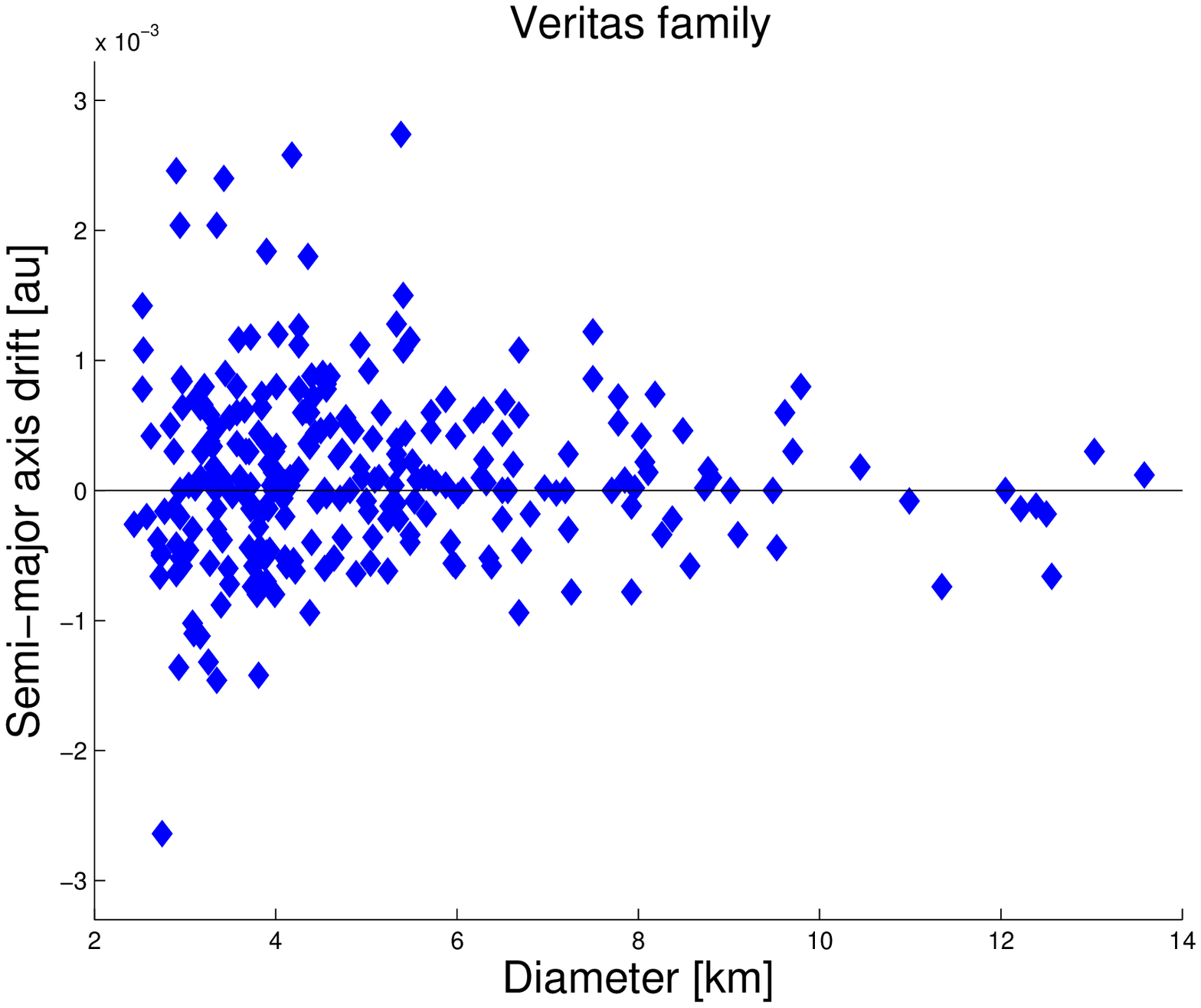}
  \caption{Semi-major axis drift ${\rm d}a$ of Veritas family members
    over the estimated age of the Veritas family (8.23 Myr). The drifts tend 
    to be smaller for larger members, as expected from the size-dependency of
    the Yarkovsky effect.}
\label{fig: drift_da}
\end{figure}

As a next step, we extracted the semi-major axis drifts ${\rm d}a$ from clones
that show the best convergence at the nominal family age. The result is shown
in Fig.~\ref{fig: drift_da}. We also obtained ${\rm d}a$ values at the limits
of the age range (7.9 and 8.6~Myr).  The distribution of ${\rm d}a$ values is
similar in all these cases.  As expected from the standard theory on the
Yarkovsky effect, the larger asteroids have smaller ${\rm d}a$ values. We
discuss this in more detail in Sects.~\ref{sec: drift_exp} and
\ref{sec: comp}. A list of the 274 studied 
asteroids with their proper elements $a,e, \sin{i}$, proper frequencies $g$ and 
$s$, Lyapunov exponents $LCE$, and estimated drift speeds ${\rm d}a/{\rm d}t$ 
is given in Table~\ref{table: Veritas_members}, available in its
full length in the electronic version of the paper.

\section{Effects of encounters with (1) Ceres and (10) Hygiea}
\label{sec: encounters_run} 

To evaluate the effect
that close encounters with massive main belt asteroids may have had on
the convergence of the secular angles, we repeated the
integration of our selected 274 Veritas members, but this time 
we also included the gravitational effect of: (i) (1) Ceres, and (ii) both 
(1) Ceres and (10) Hygiea.  The latter asteroid, the fourth most massive
in the main belt, was included because of its orbital proximity to the
Veritas family and the possible role that the $g-g_{Hygiea}$ secular 
resonance (see Appendix~1 for a discussion of the possible effect 
of this resonance).

\begin{figure}
  \centering
  \centering \includegraphics [width=0.45\textwidth]{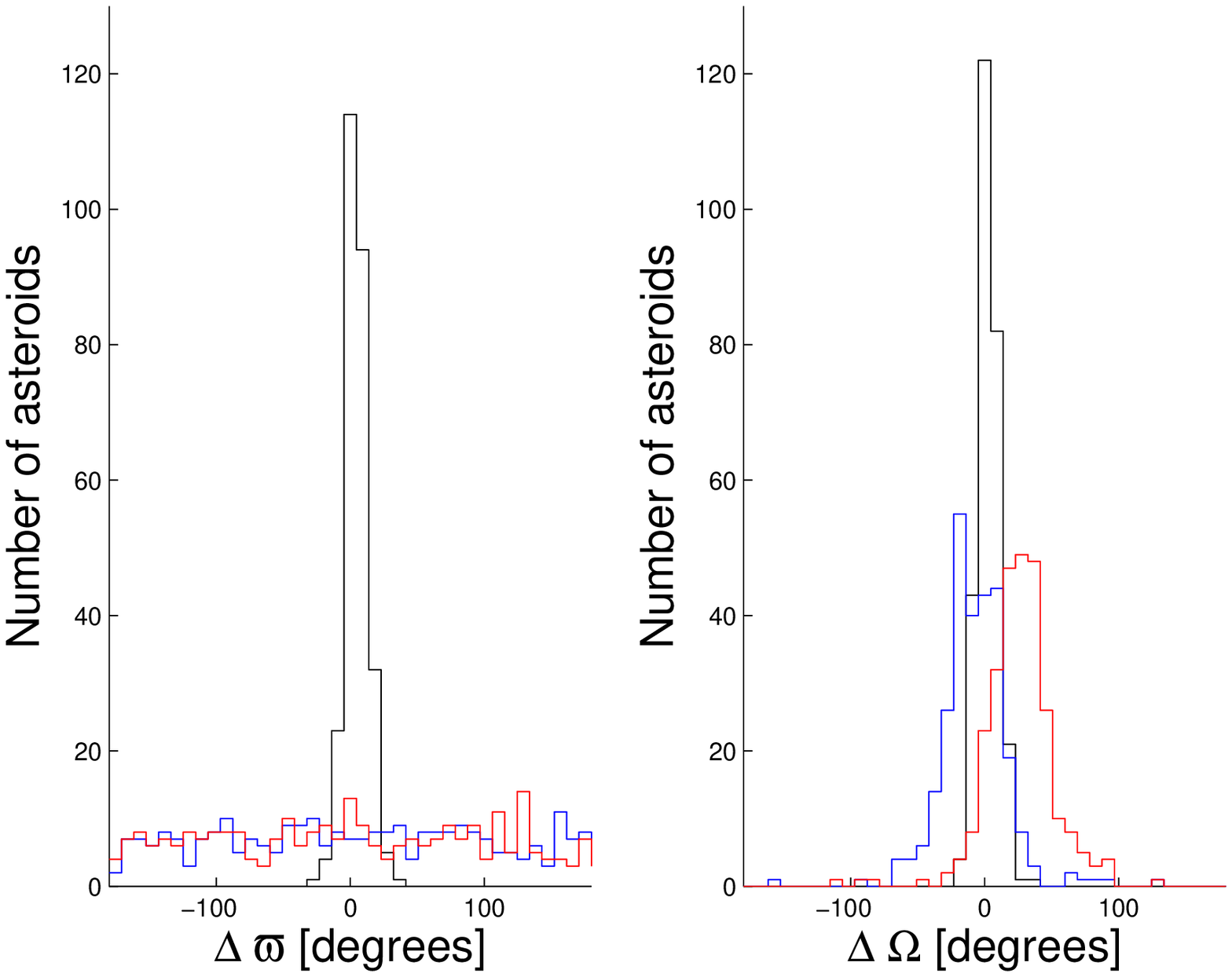}
\caption{Distribution of changes in $\varpi$ and $\Omega$ for the 274 integrated
asteroids for the case without Ceres and Hygiea (black line), with Ceres
(blue line), and with Ceres and Hygiea (red line).} 
\label{fig: ceres_enc}
\end{figure}

Fig.~\ref{fig: ceres_enc} shows the distribution of changes of $\varpi$ and
$\Omega$ in different cases. The $da$ values from the Yarkovsky effect were
kept constant in all the three cases.  In principle, fine tuning of these
values, as performed in the previous section, could compensate the
effects of encounters with Ceres and Hygiea, but, since the occurrence
of a given encounter with a massive body at a given time depends on the
Solar System model used \citep{Carruba_2012}, adding more massive bodies to
the simulations would require a different fine tuning.

Encounters with Ceres and Ceres/Hygiea increase the standard deviation of t
he distribution in $\Delta \Omega$ from $8.4^{\circ}$ up to $26.2^{\circ}$ and
$26.0^{\circ}$, respectively.
Surprisingly, they completely destroy the convergence in $\Delta \varpi$.  Both
models cause the standard deviation of the distribution in $\Delta \varpi$
to reach values near $101.3^{\circ}$, corresponding to a uniform distribution.
What is causing this remarkable behavior?

\begin{figure}
  \centering
  \centering \includegraphics [width=0.45\textwidth]{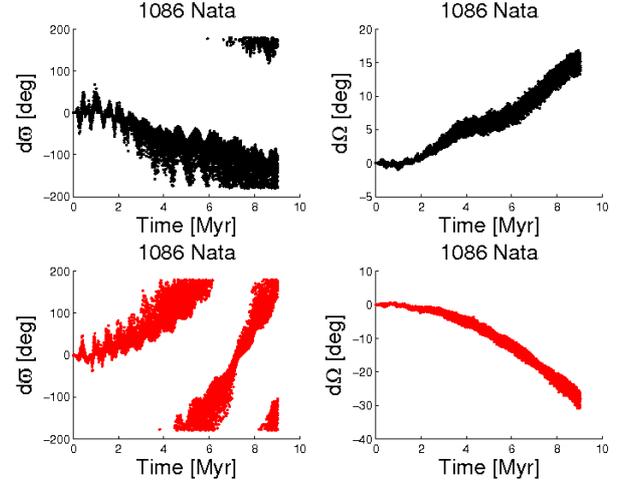}
  \caption{$\Delta \varpi$ and $\Delta \Omega$ values as a function of time for
    (1086) Nata when Ceres was considered as a massive perturber (top panels,
    black dots) and when Ceres and Hygiea were both considered as perturbers
    (bottom panels, red dots). The values were computed relative to a
    reference case where no massive asteroid perturbers were included in the
    integration.}
\label{fig: Nata_diff}
\end{figure}

Figure~\ref{fig: Nata_diff} show values of $\Delta \varpi$ and $\Delta \Omega$
values as a function of time for (1086) Nata with respect to a reference case 
where no massive asteroids were included in the integration.
It can be noted that while changes in $\Delta \Omega$ are smaller than 
$30^{\circ}$ in both models, changes in $\Delta \varpi$ are bigger than 
$180^{\circ}$ in the first model, and two complete circulations of 
this angle were observed in the second model.  We can also observe that 
the time behavior of the angles is different in the two models.  
While $\Delta \varpi$ increases in the first model, it decreases in the
second one (and vice versa for $\Delta \Omega$). 

Let us now try to understand what makes the secular angle convergence in 
the Veritas family so sensitive to gravitational perturbations of massive
asteroids, and what it implies for the realistic uncertainty in our
determination of the drift rates caused by the Yarkovsky effect. Our method is
as follows.

The massive asteroids may affect the nominal convergence of
secular angles in two ways: (i) a direct contribution to the secular
frequencies $s$ (node) and $g$ (pericenter), or (ii) an indirect
effect, which consists of perturbations of the semi-major axis $a$,
being then reflected in nodes and pericenters via dependence
of $s$ and $g$ on $a$. We believe the latter is dominant, and we
will try to demonstrate it in the case of Ceres' influence.
Obviously, the effect is larger in the longitude of pericenter just
because the gradient $\partial g/\partial a$ is nearly an order of magnitude
larger than $\partial s/\partial a$ in the Veritas family (see, e.g.,
Appendix~1).

We postulate that the nature of Ceres' perturbation in semi-major axis
of Veritas orbits is due to stochastic jumps during sufficiently close
encounters. These are favored by two facts: (i) the orbit pericenter $q$ for
Veritas orbits is close to the orbit apocenter $Q$ of Ceres, and
(ii) the mean inclination of Veritas orbits is nearly the same as the
mean inclination of Ceres' orbit. To probe how (i) and (ii) influence
circumstances of close encounters to Ceres, we used the \"Opik theory
to compute intrinsic collision probability $p_i$ of the two orbits.
Indeed, we obtained $p_i\simeq 3.5\times 10^{-17}$ km$^{-2}$~yr$^{-1}$,
which is slightly times larger than the average in the main belt
(e.g., \citet{Bottke_1994}).

Given the difference of $s$ frequencies of Veritas members and Ceres,
we note that every $\simeq 60$~kyr the orbital planes get very close to
each other. These are the moments when the instantaneous collision
probability with Ceres becomes even larger than the average stated
above. Obviously, in order for a close encounter to really happen, Ceres
must be close to its aphelion and Veritas member close to its
perihelion. The difference in $g$ frequencies is much larger than
in $s$ frequencies, so the timescale of very favorable collision
probability of Veritas members to Ceres is $\simeq 60\times (360/5)$~kyr
or $\simeq 4$~Myr (assuming, for simplicity, that in the Ceres
orbital plane the orbit of Veritas members must be oriented within $\simeq
5^\circ$ with respect to the optimum aphelion-perihelion configuration).
Therefore, during the estimated age of Veritas family, a typical
member may undergo up to three such close encounters to Ceres.
This is consistent with data in the top and right panel of Fig.~13.
Note that the additional nodal drift of (1086)~Nata underwent two
changes of rate at approximately $3$~Myr and again at about $6.5$~Myr.

Obviously, none of the currently observed Veritas members impacted Ceres
during its lifetime. We may, however, estimate an order of magnitude of
its closest approach $R_{\rm app}$. Considering again the \"Opik-theory
approach, the condition for $R_{\rm app}$ reads: $p_i R_{\rm app}^2 T
\simeq 1$, where $T=8.3$~Myr the age of the family. Using this relation we
obtain $R_{\rm app}\simeq 5.6\times 10^5$~km. This result is averaged
over all possible orbit configurations. Assuming special coplanarity
and alignment conditions would have a much larger collision probability,
we may assume the really closest approaches of Veritas fragments to Ceres
that occurred few times in the past had $R_{\rm app}\simeq 10^4$~km.

Finally, we estimate the magnitude of orbital semi-major axis
jump $\delta a$ of Veritas member during such a close encounter with
Ceres. Obviously, the result depends on details of the encounter. However,
here we are interested in obtaining the order of magnitude result only.
Assuming the change in binding energy to the Sun is of the order of
magnitude of potential energy in the Ceres gravitational field at the
moment of the closest approach, we have $(\Delta a/a)\simeq (M_{\rm 1}/M_{\rm 0})
(a/R_{\rm app})$, where $M_{\rm 0}$ and $M_{\rm 1}$ are the masses of the Sun and
Ceres and $a$ is semi-major axis of the Veritas fragment. We obtain
$\delta a \simeq 7\times 10^{-5}$~au in one encounter. If statistically
two such encounters may happen the accumulated $\Delta a$ becomes about
$10^{-4}$~au.

To tie this information to our previous results we note that the
characteristic accumulated change in semi-major axis of small
Veritas members to achieve orbital convergence is $\simeq (5-10)\times
10^{-4}$~au (see Fig.~11). While a part of this value may be due to
stochastic effects of Ceres encounters, the main effect is likely due to
other dynamical phenomenon. We favor the Yarkovsky effect and discuss
details in Secs.~7 and 8. In this respect, however, the random component
due to encounters to Ceres and other massive objects in the belt represents
a noise. Given the estimates above, its realistic level is
$\simeq 1.3\times 10^{-5}$ au~Myr$^{-1}$ in the mean drift rate due
to Ceres along. To stay realistic, we shall assume three times larger
value to account for the effects of other massive bodies such as (10)
Hygiea.

\begin{figure}[pt]
\begin{center}
 \includegraphics[width=8cm]{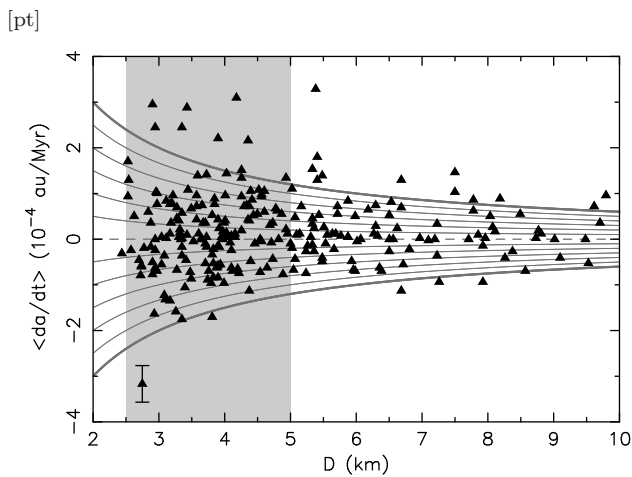}
\end{center}
\caption{Adjusted semi-major axis drift rates $da/dt$ to achieve optimum
 convergence in secular angles of 264 $D<10$~km asteroids in the R$_1$
 region of Veritas family versus their estimated size $D$. The values
 correspond to the formal best fit solution minimizing $\chi^2$ in Sect.~
(\ref{sec: meas_drift})
 at $8.23$~Myr. The thick gray lines correspond to our estimate of maximum
 values of the Yarkovsky effect for nominal physical parameters, or
 $\pm 2.4\times 10^{-4}$ au~Myr$^{-1}$ at $D=2.5$~km (see Fig.~\ref{dadt})
 and the characteristic $\propto 1/D$ size dependence. The thin gray lines
 for respectively smaller $da/dt$ values at the reference size of $2.5$~km.
 The uncertainty interval in $da/dt$ shown for the outlying data-point
 at the bottom left corner applies to all data. This is our estimated
 realistic value that takes into account: (i) formal variation of
 $da/dt$ within the sigma interval of the Veritas family age solution
 (Fig.~\ref{fig: chi2_age}, panel A), and (ii) the stochastic effect of 
 Ceres and Hygiea close
 encounters on behavior of the longitude of node and pericenter of
 asteroids in the Veritas family. The light-gray rectangle indicates
 the size range $2.5$ to $5$~km used for our analysis in
 Sec.~\ref{sec: meas_drift}.}
\label{dadt_solved}
\end{figure}

\section{Expected drift-rate in semi-major axis due to the Yarkovsky effect}
\label{sec: drift_exp}

For convenience of discussion in this Section, we express the empirical
accumulated change in semi-major axis in $\simeq 8.23$~Myr of the Veritas
family age shown in Fig.~\ref{fig: drift_da} in terms of the equivalent
mean rate. This is shown in Fig.~\ref{dadt_solved}. Here we pay attention to
the consistency of these values with the predictions from the Yarkovsky effect
theory. In particular, we estimate the expected values of to the Yarkovsky
effect for these bodies and compare them with the empirical values required
by orbital convergence discussed in Sec.~\ref{sec: meas_drift}.
A general information about the theory of the Yarkovsky effect may
be found in \citet{Bottke_2006} or \citet{Vokrouhlicky_2015}. Here
we only summarize few facts needed for our application.

While in principle there is only one Yarkovsky effect acting on the body 
as such, it is customary to divide it into diurnal and seasonal components.
Recall that the thermal response of the body surface on solar heating
is frequency-dependent and there are typically two primary frequencies
involved (unless tumbling rotation state that we disregard in this
study): (i) rotation frequency $\omega$, and (ii) revolution frequency 
(or mean motion) $n$ about the Sun. In virtually all asteroid applications 
$\omega\gg n$. In this limit, we may consider the thermal effects related 
to $\omega$ (the diurnal component) and those related to $n$ (the seasonal 
component) frequencies separately. Their mutual interaction is negligible 
(see, e.g., \citet{Vokrouhlicky_1999}).
In applications of the Yarkovsky effect to the motion of small asteroids
only the diurnal component has been considered so far, while the seasonal 
component has been neglected. There are two reasons that justified this
approach. First, in the limit of very low-enough surface thermal inertia 
values, appropriate for multi-kilometer and larger asteroids in the main 
belt whose surfaces are expected to be covered with fine regolith layer, 
the magnitude of the seasonal effect becomes
$\simeq \sqrt{n/\omega}$ smaller than that of the diurnal effect.
Second, the seasonal effect may become important only for a restricted
interval of obliquity values near $90^\circ$. For evolved-enough populations
of asteroids, the Yarkovsky-O'Keefe-Radzievskii-Paddack (YORP) effect
makes the obliquity pushed towards the extreme values $0^\circ$ or
$180^\circ$ (e.g., \citet{Bottke_2006} or \citet{Vokrouhlicky_2015})
where the seasonal Yarkovsky effect becomes nil. However, in the case
of Veritas fragments, none of these arguments might be satisfied. Thermal
inertia of fresh, C-type objects might be anomalously large, compared
to the data of all asteroids. In the same time, the YORP effect likely
did not have enough time to push the obliquity values of the currently
observed Veritas members to their asymptotic values (see also 
Sec.~\ref{sec: YORP}).  As a result, we include both the diurnal and 
the seasonal components of the Yarkovsky effect in our analysis. 

Given only the moderate value of the orbital eccentricity of the Veritas 
family members, we restrict to the Yarkovsky model applicable 
to circular orbits. This is sufficient, because the eccentricity 
corrections to the mean orbital change in semi-major axis are only of the 
second order in $e$. At the same time, we limit ourselves to the 
analytic model for the spherical shape of the asteroids and linearized 
boundary conditions of the heat conduction. While approximate, the main
advantage is that we thus dispose of simple analytic formulas for the
secular long-term change of the orbital semi-major axis. Finally, we note 
that the penetration depth $\ell$ of the thermal wave for both diurnal
and seasonal effects is much smaller than the characteristic radius
$R$ of the known Veritas members ($\ell$ being at maximum few meters
even for the seasonal effect; e.g., \citet{Bottke_2006, Vokrouhlicky_2015}). 
In this case, we neglect corrections of the order
$\propto \ell/R$ or higher in our analysis.

Using all these approximations, we find that the mean semi-major
axis drift-rate due to the diurnal variant of the Yarkovsky effect is 
given by
\begin{equation}
 \left(\frac{da}{dt}\right)_{\rm diu} \simeq \frac{4\alpha}{9}\frac{\Phi}{n}
  \frac{\Theta}{1+\Theta+\frac{1}{2}\Theta^2 \rule{0pt}{2.3ex}}
  \cos\gamma\;, 
\label{dadt_diu}
\end{equation}

\noindent
where $\alpha=1-A$, with $A$ the Bond albedo, $\Phi=(\pi D^2 F)/
(4mc)$, with the size (diameter) $D$ of the body, $F\simeq 136.3$~W/m$^2$
the solar radiation flux at the mean heliocentric distance of the Veritas
family, $m$ the mass of the body, $c$ is the velocity of light,
and $n$ the orbital mean motion. Note that $\Phi\propto 1/D$, which 
implies the Yarkovsky effect magnitude is inversely proportional to the 
asteroid size. Therefore it is negligible for large Veritas members such 
as (1086)~Nata, but becomes important as soon as $D$ becomes smaller than 
$\simeq 5-10$~km. Similarly, denoting $\rho$ the bulk density of the 
asteroid, one has $\Phi\propto 1/\rho$, again the inverse proportional
scaling with this parameter. The thermal 
parameter $\Theta = \Gamma\sqrt{\omega}/(\epsilon\sigma T_\star^3)$
depends on of the surface thermal inertia $\Gamma$, the rotation
frequency $\omega$, the surface infrared emissivity $\epsilon$,
the Stefan-Boltzmann constant $\sigma$ and subsolar temperature
$T_\star=[\alpha F/(\epsilon\sigma)]^{1/4}$. Finally, $\gamma$
is the obliquity of the asteroid spin axis.


\begin{figure*}
  \centering
  \begin{minipage}[c]{0.49\textwidth}
    \centering \includegraphics[width=3.1in]{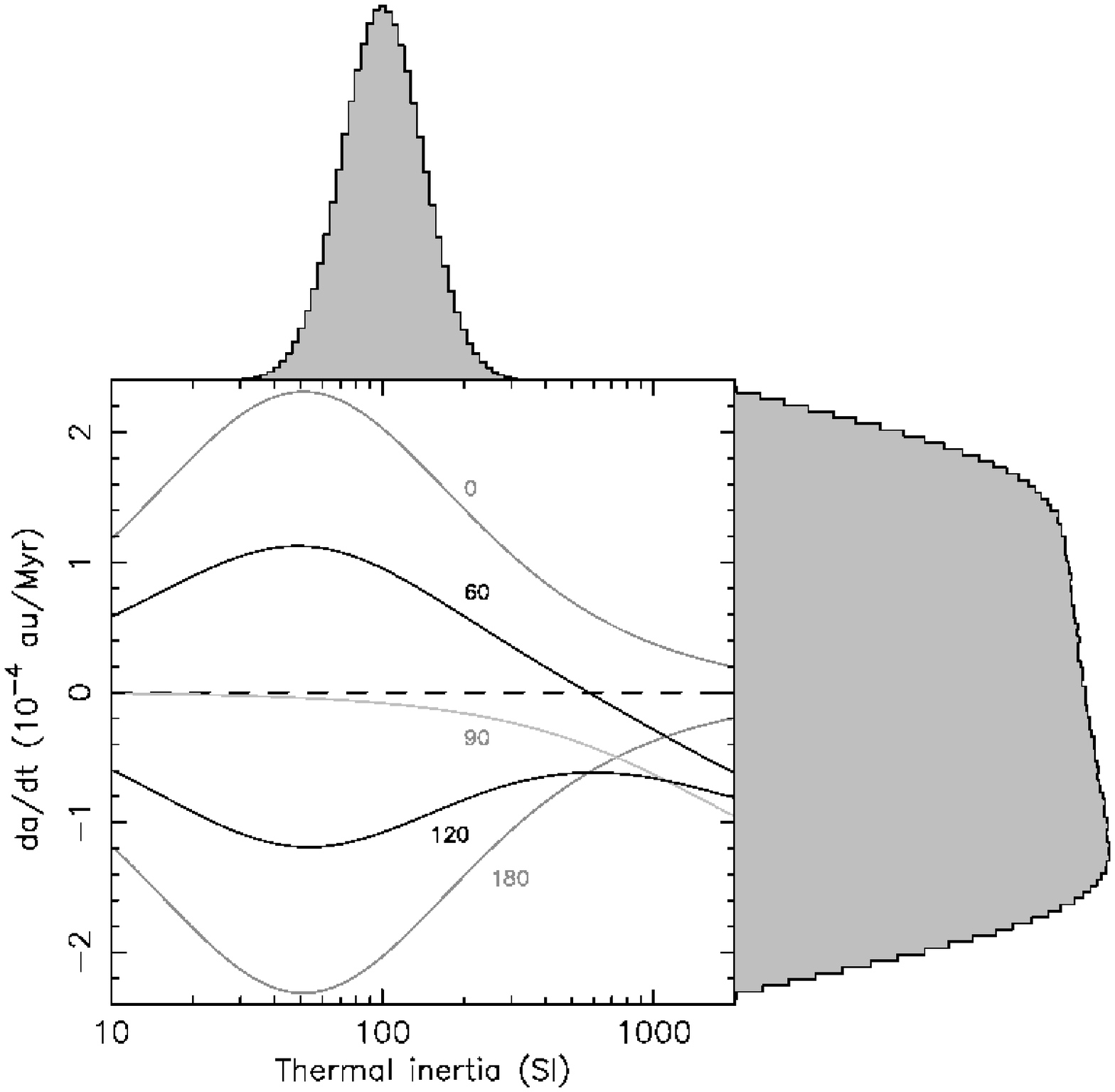}
  \end{minipage}%
  \begin{minipage}[c]{0.49\textwidth}
    \centering \includegraphics[width=3.1in]{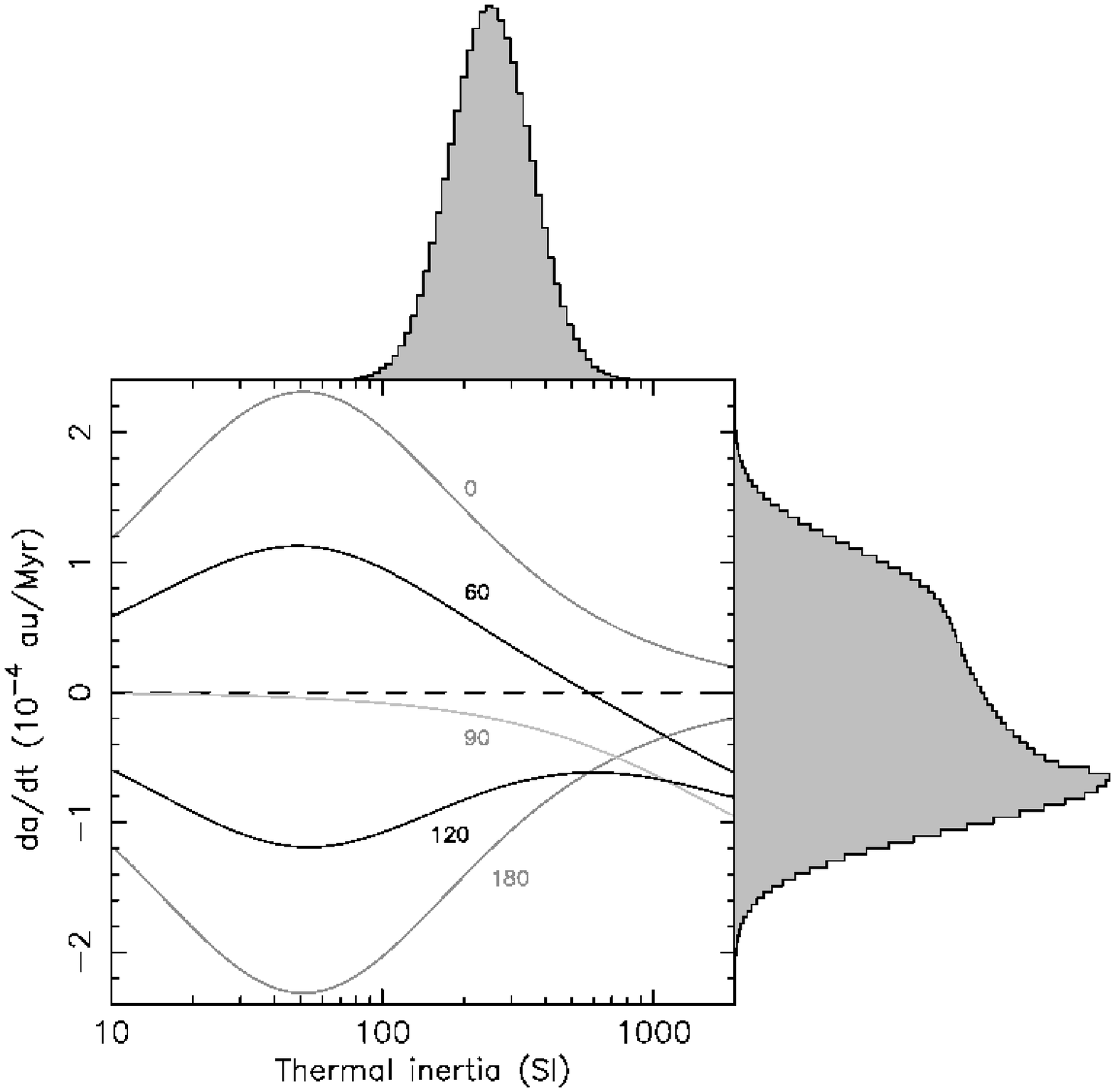}
  \end{minipage}
\caption{The middle part on the left and right panels shows
 dependence of the total semi-major axis drift-rate due to the
 Yarkovsky effect on the surface thermal inertia $\Gamma$ (we
 assumed $D=2.5$~km size member in the Veritas family with a
 rotation period of $6$~hr and a bulk density $\rho=1.3$
 g~cm$^{-3}$). Different curves for five representative
 values of the obliquity $\gamma=0^\circ$, $60^\circ$, 
 $90^\circ$, $120^\circ$ and $180^\circ$ (labels). Assuming a population
 of bodies with a log-normal distribution of $\Gamma$ values shown
 by the top histograms, and isotropic distribution of spin axis
 orientations in space, one would obtain distributions of the
 $da/dt$ values shown on the right side of each of the panels.
 The difference between left and right is in the mean value of
 $\Gamma$: $100$~SI units on the left and $250$~SI units on the
 right.}
\label{dadt}
\end{figure*}

Using the same notation as above, the mean semi-major axis drift-rate
due to the seasonal component of the Yarkovsky effect reads
\begin{equation}
 \left(\frac{da}{dt}\right)_{\rm sea} \simeq -\frac{2\alpha}{9}\frac{\Phi}{n}
  \frac{{\bar \Theta}}{1+{\bar \Theta}+\frac{1}{2}{\bar \Theta}^2
 \rule{0pt}{2.3ex}} \sin^2\gamma\;. 
\label{dadt_sea}
\end{equation}

\noindent
This is very similar to (\ref{dadt_diu}), except for the diurnal thermal 
parameter $\Theta$ now replaced with its seasonal counterpart ${\bar \Theta}
=\Gamma\sqrt{n}/(\epsilon\sigma T_\star^3)$. Note that the only difference 
consists in the rotation frequency $\omega$ being substituted with the 
orbital mean motion $n$. Thus ${\bar \Theta}$ is always smaller than
$\Theta$ since ${\bar \Theta}/\Theta = \sqrt{n/\omega}$. Importantly-enough,
the diurnal and seasonal effects have also different dependence on the
rotation pole obliquity $\gamma$, the former being maximum at
$\gamma=0^\circ$ and $180^\circ$, the latter at $\gamma=90^\circ$.
Hence, the aforementioned YORP-driven depletion of the obliquity 
distribution at mid-$\gamma$ values contributes to dominance of 
the diurnal effect. However, when the YORP effect does not have enough time
to modify obliquity values, as we check in the next section, there is
no obvious reason to neglect the role of $\gamma\simeq 90^\circ$
obliquity values.

The total rate in semi-major axis is simply $da/dt=(da/dt)_{\rm diu}+
(da/dt)_{\rm sea}$. For a given body, $da/dt$ depends on a number
of physical parameters described above. In some cases the dependence 
conforms to a simple scaling. Thus, $da/dt\propto \Phi\propto 1/(D\rho)$.
For other parameters the dependence is less obvious and needs to
be explored numerically. This is the case of the surface thermal
inertia $\Gamma$, rotation frequency $\omega$ and obliquity 
$\gamma$. Obviously, we do not know any of these values for 
small members in the Veritas family. It is not, however, our intent here to
speculate about individual bodies in the Veritas family. Rather,
by comparison with the derived $da/dt$ values in 
Sec.~\ref{sec: meas_drift}, we 
may be only able to say something about distribution of these
parameters in the whole sample of small Veritas members.

To assist with this goal, we performed the following numerical experiment
which may serve as a template for comparison with real Veritas data.
Figure~\ref{dadt} shows results where we fixed a reference asteroid
size $D = 2.5$~km, bulk density $\rho=1.3$ g~cm$^{-3}$ and rotation
period $6$~hr (implying thus $\omega\simeq 2.9\times 10^{-4}$ 
rad~s$^{-1}$). The chosen size corresponds to that of the smallest
Veritas members, for which we were able to determine $da/dt$ value
from the secular angles convergence (Fig.~\ref{dadt_solved}). The assumed bulk
density is the best guess for the C-type taxonomy of the Veritas
family (e.g., \citet{Scheeres_2015}). Two more parameters remain
to be selected for predicting the Yarkovsky value of $da/dt$,
namely the surface thermal inertia $\Gamma$ and obliquity
$\gamma$. For the latter, we assume an isotropic distribution
of the spin axes in space. This is, in fact, the underlying 
hypothesis that we are going to test in comparison with the data. For
the former, we assumed a log-normal distribution with two
different mean values of $100$~SI units (left panel) and $250$~SI 
units (right panel). Both values are within the expected range
of thermal inertia values of small, C-type asteroids (Fig.~9 of
\citet{Delbo_2015}).

The maximum range of possible $da/dt$ values due to the Yarkovsky
effect spans, for the chosen parameters, from about $-2.4\times 10^{-4}$
au~Myr$^{-1}$ to $2.4\times 10^{-4}$ au~Myr$^{-1}$. Given the $6$~hr
rotation period it corresponds to $0^\circ$ or $180^\circ$ obliquity
values and $\Gamma\simeq 50$ in SI units. These limits would be
the same for other rotation periods, because the diurnal Yarkovsky
effect, the only component contributing at extreme obliquity values,
is invariant when $\Gamma$ and $\omega$ change preserving $\Gamma\sqrt{\omega}$
value (see Eq.~\ref{dadt_diu}). So for longer rotation periods the
optimum thermal inertia would just slightly shift to the larger values.
These maximum values favorably compare with majority of our solved
$da/dt$ values for small fragments in the Veritas family 
(Fig.~\ref{dadt_solved}). The few outliers beyond these limits, may be 
bodies residing on
slightly unstable orbits for which our analysis is not so accurate
(for instance due to close encounters to Ceres or Hygiea, or secular
effects due to these massive bodies in the main belt).
Alternately, these cases may also correspond to anomalously
low-density fragments, or bodies for which the absolute magnitude has 
not been determined accurately and they actually have quite smaller size.

The Yarkovsky effect at $90^\circ$ obliquity is the pure seasonal
contribution. At low $\Gamma$ values this component is basically
nil, but at $\Gamma\simeq 700$ in SI units it rivals the maximum value
of the diurnal effect. At still reasonable $\Gamma$ values between
$300$ and $400$ in SI units its contribution is not 
negligible and needs to be taken into account. The contribution of
the seasonal effect makes the total Yarkovsky $da/dt$ value at generic
obliquities, such as $60^\circ$ or $120^\circ$ on Fig.~\ref{dadt},
bent toward negative values for sufficiently large thermal
inertia, breaking thus the symmetry between the positive and negative
$da/dt$ values.

Let us assume a population of Veritas fragments with an isotropic 
distribution of spin axes and log-normal tight distributions of 
$\Gamma$ values shown by the top histograms on the left and right 
panels of Fig.~\ref{dadt}. The histograms along the right ordinate
then indicate what would be the distribution of $da/dt$ values
for this sample. As already expected, the contribution of the
seasonal Yarkovsky effect makes it that in both cases the
negative $da/dt$ values are more likely. However, for low thermal
inertia values (left panel), the asymmetry is only small,
$52.4$\% vs $47.6$\% of all cases only. This asymmetry becomes
more pronounced for larger thermal inertia values (right panel),
for which $59.2$\% cases have $da/dt$ negative and only
$40.8$\% cases have $da/dt$ positive. Playing with more assumptions
about the thermal inertia distributions one may thus create other
templates for the $da/dt$ distributions for samples of fragments
that have their spin axes isotropically distributed in space.

Obviously, if none of such templates corresponds to the observed
distribution of $da/dt$ values in the Veritas family, the
underlying assumption of isotropy of the rotation axes must be
violated. With only limited data, and a lack of information 
about fragments other physical parameters, we cannot solve
for the spin axis distribution, but at least say some trends.
This is the goal of our analysis in Sec.~\ref{sec: YORP}.

\subsection{A possible role of the YORP effect?}
\label{sec: YORP}

Before we proceed with comparison of the observed and modeled
$da/dt$ values, we pay a brief attention to the neglected role
of the YORP effect. In particular, we assumed that the $\simeq 8.3$~Myr
age of the Veritas family is short enough that the YORP toques
did not result in a significant change of obliquity $\gamma$
of its members (such that $\gamma$ could be considered constant
in our analysis). This may look odd at the first sight, because
the Yarkovsky and YORP effects are just two faces of one physical
phenomenon, namely the recoil effect of thermally re-radiated sunlight
by the asteroid surface. 

The key element in understanding this issue is a different dependence
on the size $D$. While the Yarkovsky effect is only inversely
proportional to $D$ ($da/dt \propto 1/D$, see above), the YORP effect
is inversely proportional to $D^2$. As a result, its importance
decreases much faster with increasing size of the bodies.

The model predictions of the YORP effect are much less certain than
those of the Yarkovsky effect (see, e.g., discussion in
\citet{Vokrouhlicky_2015}). In this situation we rather use real data from the
Karin family \citep{Carruba_2016a}, and re-scale them to the
case of the Veritas family. This simple re-scaling procedure must take
into account different heliocentric distances ($a\simeq 2.86$~au vs
$a\simeq 3.16$~au), bulk densities ($\rho\simeq 2.5$ g~cm$^{-3}$ vs 
$\rho\simeq 1.3$ g~cm$^{-3}$) and ages ($T\simeq 5.8$~Myr vs $T\simeq
8.3$~Myr). The accumulated change $\Delta \gamma$ in obliquity
scales with $\propto T/(\rho a^2 D^2)$. This makes us to conclude
that $D\simeq 2$~km Karin asteroids should accumulate about the same
obliquity change as $D\simeq 3$~km Veritas asteroids (the smaller
bulk density and larger age playing the main role). Looking at
Fig.~5 of \citet{Carruba_2016a}, we conclude that the YORP effect
is negligible for those sizes (as it has been also verified in that
paper). Since $D\simeq 3$~km are about the smallest asteroids we 
presently observe in the Veritas family, the omission
of the YORP effect in this study is justified. For sake of
comparison we note, that the population of $D\simeq 1.4$~km Karin 
asteroids already revealed traces of the YORP effect (Fig.~5 in
\citet{Carruba_2016a}). Therefore we predict that, when the future 
observations will allow to discover $D\simeq 2$~km members in the
Veritas family, we should start seeing similar YORP pattern
as in the Karin family (i.e., depletion of the $da/dt\simeq 0$
values).

\begin{figure}
\centering
\centering \includegraphics [width=0.45\textwidth]{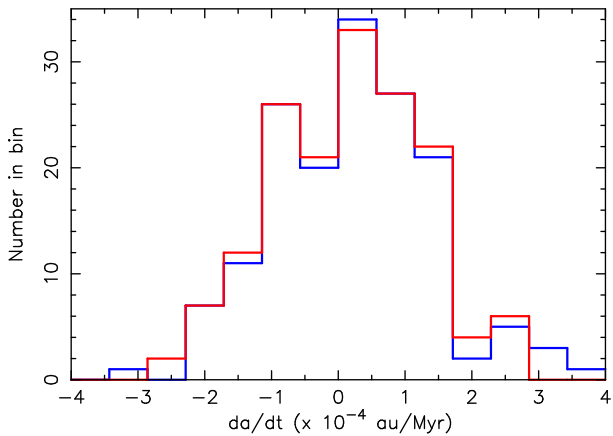}
\caption{Result of a formal fit of the observed $da/dt$ values
 for $160$ Veritas members with estimated size $\leq 5$~km all
 mapped to a reference size of $2.5$~km using the $\propto 1/D$
 scaling (blue histogram). The bin width, $\simeq 0.6\times 10^{-4}$
 au~Myr$^{-1}$ , is only slightly smaller than the estimated
 realistic uncertainty of the $da/dt$ values of individual asteroids
 because of the effect of massive bodies in the main belt. The red
 histogram is a best-fit from a simulation, where we assumed (i)
 isotropic distribution of rotation poles, (ii) rotation
 period $7$~hr, (iii) mean surface thermal inertia $158$ in SI
 units, and (iv) bulk density $1.07$ g~cm$^{-3}$.}
\label{dadt_fit}
\end{figure}

\section{Comparison of the observed and modeled Yarkovsky drift
 values}
\label{sec: comp}

Here we describe the sample of the solved-for values of the Yarkovsky 
drift $da/dt$ from Sec.~\ref{sec: meas_drift} in some detail. However, 
the reader is to be warned upfront that we show here just an example of 
many possible solutions. The data are simply not constraining the model 
enough at this moment.

We already observed the overall consistency of the maximum determined
$da/dt$ with the Yarkovsky prediction for the sizes of Veritas
members. Here we consider a sub-sample of 190 asteroids with size
$D\leq 5$~km. This is because data for small enough asteroids
should enable a better description of the Yarkovsky effect and be
less contaminated in a relative sense by the perturbation from
Ceres and massive asteroids.   We use the inverse-proportional relation between
$da/dt$ and $D$ and map all these values to the reference size
$D_{\rm ref}=2.5$~km. This corresponds to the smallest-observed Veritas
members (Fig.~\ref{dadt_solved}). The blue histogram at Fig.~\ref{dadt_fit} 
shows distribution of these mapped values. We used $\simeq 0.6 \times 10^{-4}$
au~Myr$^{-1}$
width of the bins for two reasons: (i) the number of asteroids in
the sample is not very large, and (ii) the realistic uncertainty of the
individual $da/dt$ solutions is $\simeq \pm 0.4\times 10^{-4}$ au~Myr$^{-1}$,
only slightly larger than the bin. The latter is mainly due to the stochastic 
effect of close encounters with Ceres and Hygiea 
(Sec.~\ref{sec: encounters_run}). Therefore 
it does not make sense to use smaller bin-size. There are 95 objects with 
$da/dt>0$, and only 65 with $da/dt<0$, in our sample. At the first sight,
this precludes a possibility of isotropic distribution of spin axes
(see discussion in Sec.~\ref{sec: drift_exp}). However, we demonstrate below 
that the small statistics of bodies still allows to match the data with the
underlying pole isotropy. Obviously, solutions with anisotropic pole
distribution are also possible, but these are not attempted here,
because they would involve many more unconstrained degrees of freedom
in the model.

\begin{figure}[pt]
\begin{center}
 \includegraphics[width=8cm]{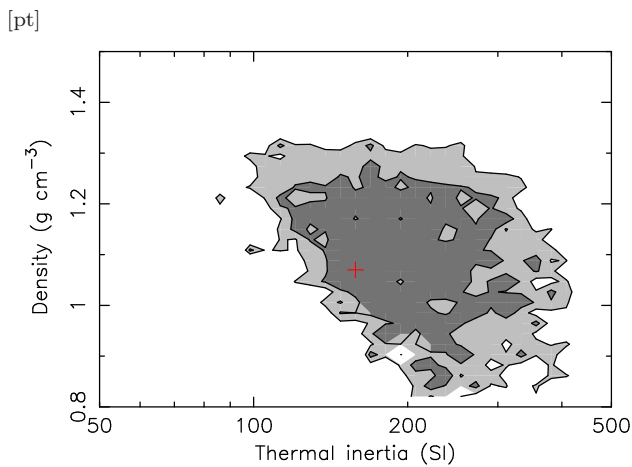}
\end{center}
\caption{Region of admissible $\chi^2$ values in the parametric
 space surface thermal inertia ${\bar \Gamma}$ (abscissa) and
 bulk density $\rho$ (ordinate) of the Veritas-family members.
 The $\chi^2$ is a measure of a success to match distribution of the
 Yarkovsky rate of change of the semimajor axis $da/dt$ obtained
 for $160$ $D\leq 5$~km asteroids in the R$_1$ region of the
 family (see the text). At each grid-point in the plane we also
 tested rotation periods between $4$~hr and $20$~hr and we run
 $10^4$ trials of isotropically distributed spin axes of the
 fragments. The light gray area shows where $\chi^2\leq 40$,
 the dark gray area shows where $\chi^2\leq 30$. The formally
 best-fit value $\chi^2=13$, shown by red cross, corresponds to
 ${\bar \Gamma}=158$ (SI units), $\rho=1.07$ g~cm$^{-3}$, and
 $P=7$~hr (see Fig.~\ref{dadt_fit}).}
\label{dadt_chi2}
\end{figure}

We conducted the following simple numerical experiment. Considering
160 $D=D_{\rm ref}=2.5$~km size members of the Veritas family, we 
randomly sampled a three-dimensional parametric space of (i) 
characteristic rotation period $P$ values (given the same to all bodies),
(ii) mean surface
thermal conductivity ${\bar \Gamma}$ with a tight log-normal distribution
with standard deviation of $0.1$ in ${\rm log}\,\Gamma$ (see the top
histograms on Fig.~\ref{dadt}, and (iii) bulk density $\rho$. The
tested interval of values were: (i) $4$~hr to $20$~hr for $P$, (ii)
$50$ to $1000$ in SI units for ${\bar \Gamma}$, and (iii) $0.8$ to
$1.5$ g~cm$^{-3}$ for $\rho$. We performed $25000$ trials of a random
sampling of this parametric space. For each these choices, we then 
let the code test $10^4$ random variants of spin axes obliquities. 
Altogether, we thus ran $2.5\times 10^8$ simulations. In each of them we
computed the distribution of $da/dt$ values for the sample using
Eqs.~(\ref{dadt_diu}) and (\ref{dadt_sea}).
We evaluated the formal $\chi^2=\sum (N_{\rm c}-N_{\rm o})^2$ measure
of the difference between the data $N_{\rm o}$ and the model
$N_{\rm c}$ using the bins shown in Fig.~\ref{dadt_fit}.
In particular, we considered only the ten central bins
around zero up to $da/dt$ values of $\pm 2.5\times 10^{-4}$
au~Myr$^{-1}$, because these are maximum expected Yarkovsky
values (e.g., Fig.~\ref{dadt}). The solution is largely
degenerate, since many parameter combinations resulted in
reasonable fits, so we restrict here to summarize just the main
trends. These are best shown by a projection of the $\chi^2$
values onto the plane of surface thermal inertia ${\bar \Gamma}$
and bulk density $\rho$ (Fig.~\ref{dadt_chi2}).

We note that the reasonable matches required: (i) ${\bar \Gamma}$ smaller
than $\simeq 400$ SI units, and most often being in the interval
between $150$ and $250$, (ii) rotation period values $P$ are not
constrained (thus no shown in the Fig.~\ref{dadt_chi2}), and (iii)
the density $\rho$ should be smaller than $1.3$ g~cm$^{-3}$. The
reason for (i) is explained in Sec.~7: larger thermal conductivity would
necessarily prefer negative values of $da/dt$ which is not observed in
our data. Actually the predominance of the positive $da/dt$ is achieved
only as a result of statistical noise on low-number of data in the
individual bins. The low densities are required to match the observed
minimum to maximum range of $da/dt$ values.

\section{Conclusions}
\label{sec: conc}

Our results can be summarized as follows:

\begin{itemize}

\item We identified and studied the members of the
  Veritas family and asteroids in the family background.  As in
  \citet{Tsiganis_2007}, we found that the chaotic orbits
  with Lyapunov times $<10^{5}$ yr are in the 3+3-2
  and 5-2-2 three body resonances.   (490) Veritas itself is
  currently in the 5-2-2 resonance and is characterized by
  a very short Lyapunov time.  With a possible exception
  of the $g-g_{Hygiea}$ secular resonance, secular dynamics 
  plays only a minor role in the region of the Veritas family

\item We studied physical properties of asteroids in the Veritas
  family region.   The Veritas family is mostly made of C-type
  objects of low albedo.  The mean albedo value of the Veritas
  family is 0.07.  Among Veritas members, only (490) Veritas and
  (1086) Nata have masses larger than $10^{17}$~kg.  The mass
  distribution of the family is consistent with the outcome
  of a fragmentation event.

\item We investigated the past convergence of nodal longitudes of 
  members of the Veritas family with Lyapunov times
  $>3 \times 10^{4}$ yr and refined the list of secure Veritas family 
  members. 704 asteroids have their nodal longitude converging to 
  within $\pm 60^{\circ}$ to that of (1086) Nata between 8.1 and 8.5 
  Myr ago.

\item  By performing two sets of numerical integrations with the 
  Yarkovsky force, we found that the inclusion of the Yarkovsky effect 
  is crucial for improving the convergence. The convergence of nodal 
  longitudes of 274 most regular members of the Veritas family is almost ideal.
  The best age estimate is $8.23^{+0.37}_{-0.31}$~Myr.  

\item For the first time, we were able to demonstrate the possibility of 
  convergence of the perihelion longitudes. 
  Regrettably, the effect of close encounters with Ceres and other massive 
  main belt asteroids destroys the convergence in $\varpi$, and also 
  defocuses the convergence of $\Omega$. This limits the accuracy of our 
  Yarkovsky drift estimates to $\pm 4 \times 10^{-5}$~au/Myr. 
  
\item To within this precision, we were able to obtain ${\rm d}a/{\rm d}t$
  drift rates for 274 members of the Veritas family. The drift rates are
  larger for smaller asteroids, as expected from the Yarkovsky effect. The
  inferred distribution of ${\rm d}a/{\rm d}t$ values is consistent with a
  population of objects with low-densities and low thermal conductivities. 

\item The effects of YORP cannot be discerned in the Veritas family. They 
  should become apparent when future observations will help to characterize 
  the family members with $D < 2$ km. 
   
\end{itemize}
In summary, despite the very complex dynamical environment of the Veritas
family, we were able to to refine the family age estimate, obtain Yarkovsky
drift rates for 274 family members, and constrain several key parameters such
as their bulk density and thermal conductivity. Results are consistent 
with the standard theory of the Yarkovsky effect.

\section*{Acknowledgments}
\label{sec: ack}

We are grateful to the reviewer of this paper, Prof. Andrea Milani,
for comments and suggestions that significantly improved the quality
of this work. We would like to thank the S\~{a}o Paulo State Science Foundation 
(FAPESP) that supported this work via the grant 16/04476-8, and the
Brazilian National Research Council (CNPq, grant 305453/2011-4).
VC was a visiting scientist at the Southwest Research Institute (SWRI)
during the preparation of this paper. DV's work was funded by the
Czech Science Foundation through the grant GA13-01308S. DN's work was 
supported by the NASA SSW program.
This publication makes use of data products from the Wide-field 
Infrared Survey Explorer (WISE) and Near-Earth Objects (NEOWISE), which are a
joint project 
of the University of California, Los Angeles, and the Jet Propulsion 
Laboratory/California Institute of Technology, funded by the National 
Aeronautics and Space Administration.  

\section{Appendix 1: Proper frequencies behavior in the Veritas region}
\label{sec: appendix_dsda}

\begin{figure*}
  \centering
  \begin{minipage}[c]{0.49\textwidth}
    \centering \includegraphics[width=3.1in]{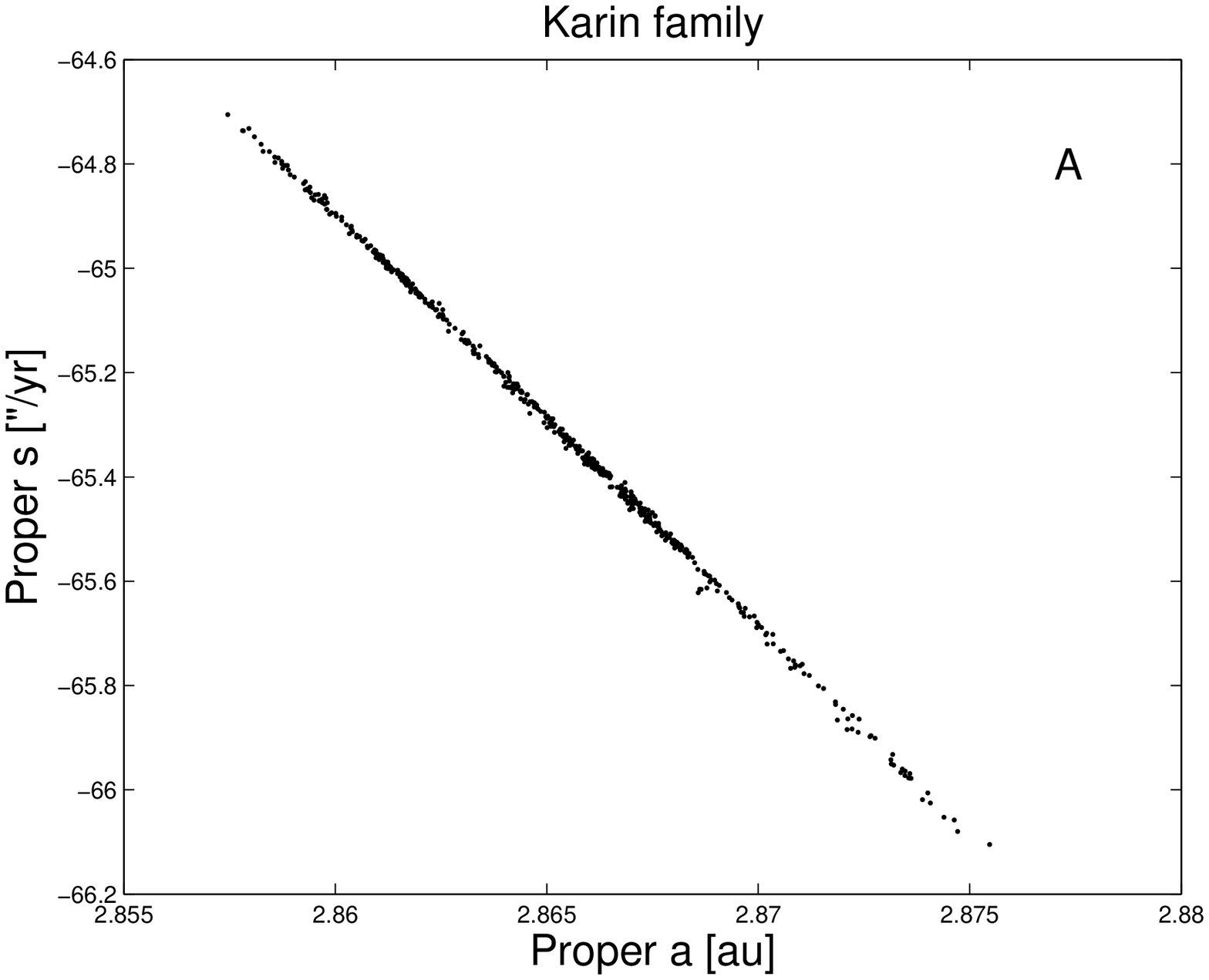}
  \end{minipage}%
  \begin{minipage}[c]{0.49\textwidth}
    \centering \includegraphics[width=3.1in]{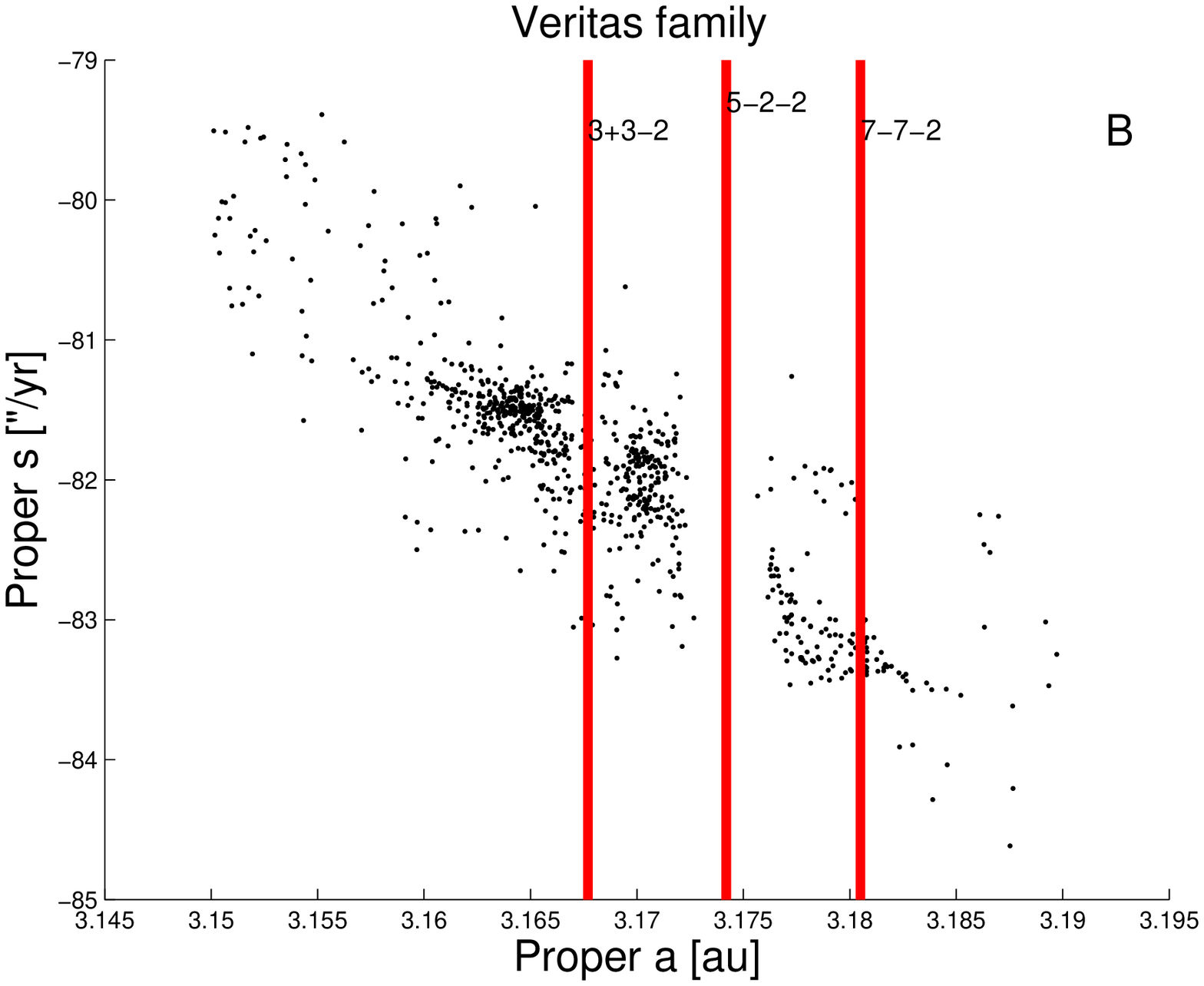}
  \end{minipage}
  \caption{An $(a,s)$ projection of members of the Karin cluster (panel A)
    and of the Veritas family (panel B).  While the $(a,s)$ distribution of
    the Karin cluster members mostly follows a single line of constant
    $\delta s/ \delta a$, the same distribution for the Veritas family
    is much more dispersed.}
\label{fig: dsa}
\end{figure*}

\begin{figure*}
  \centering
  \begin{minipage}[c]{0.49\textwidth}
    \centering \includegraphics[width=3.1in]{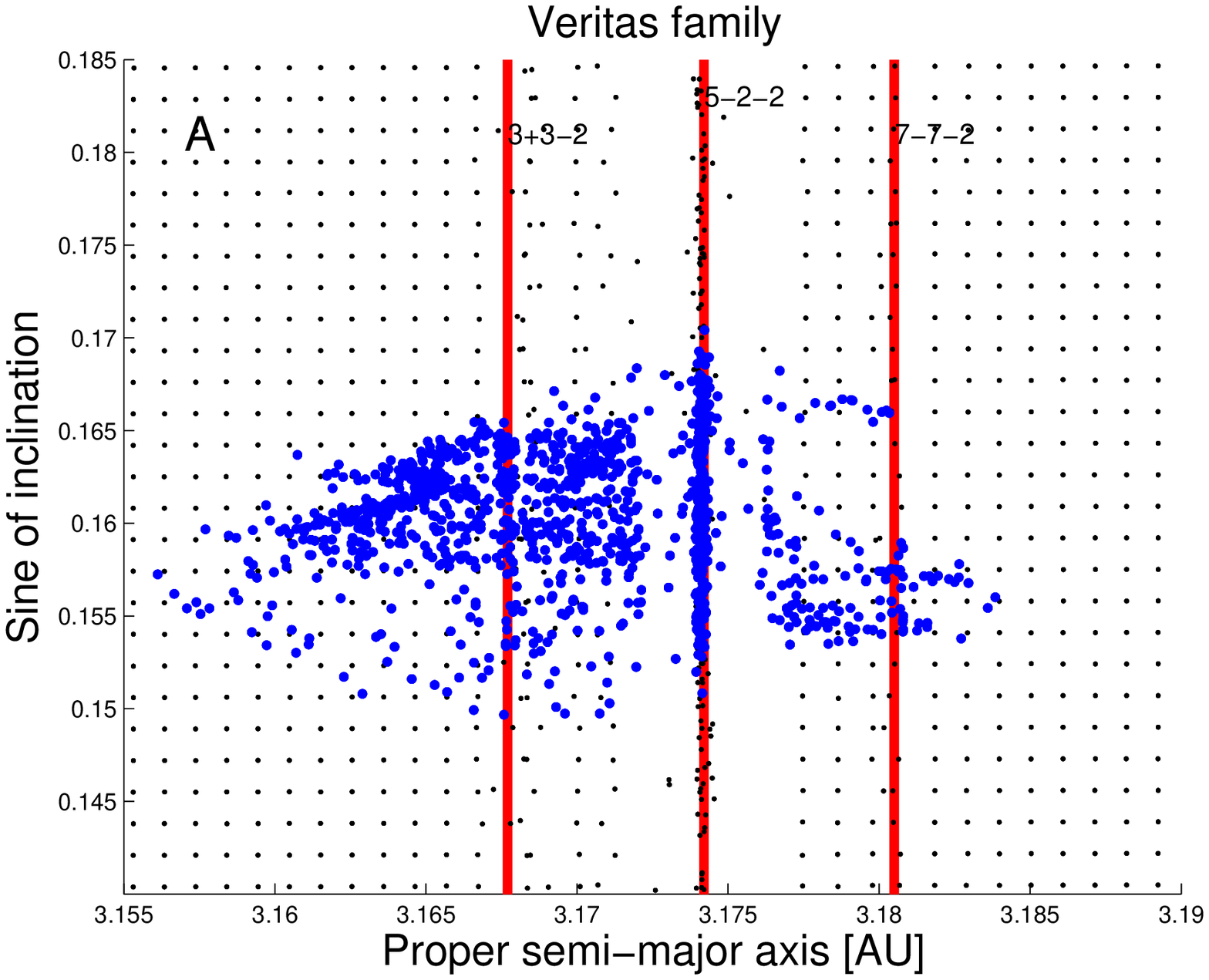}
  \end{minipage}%
  \begin{minipage}[c]{0.49\textwidth}
    \centering \includegraphics[width=3.1in]{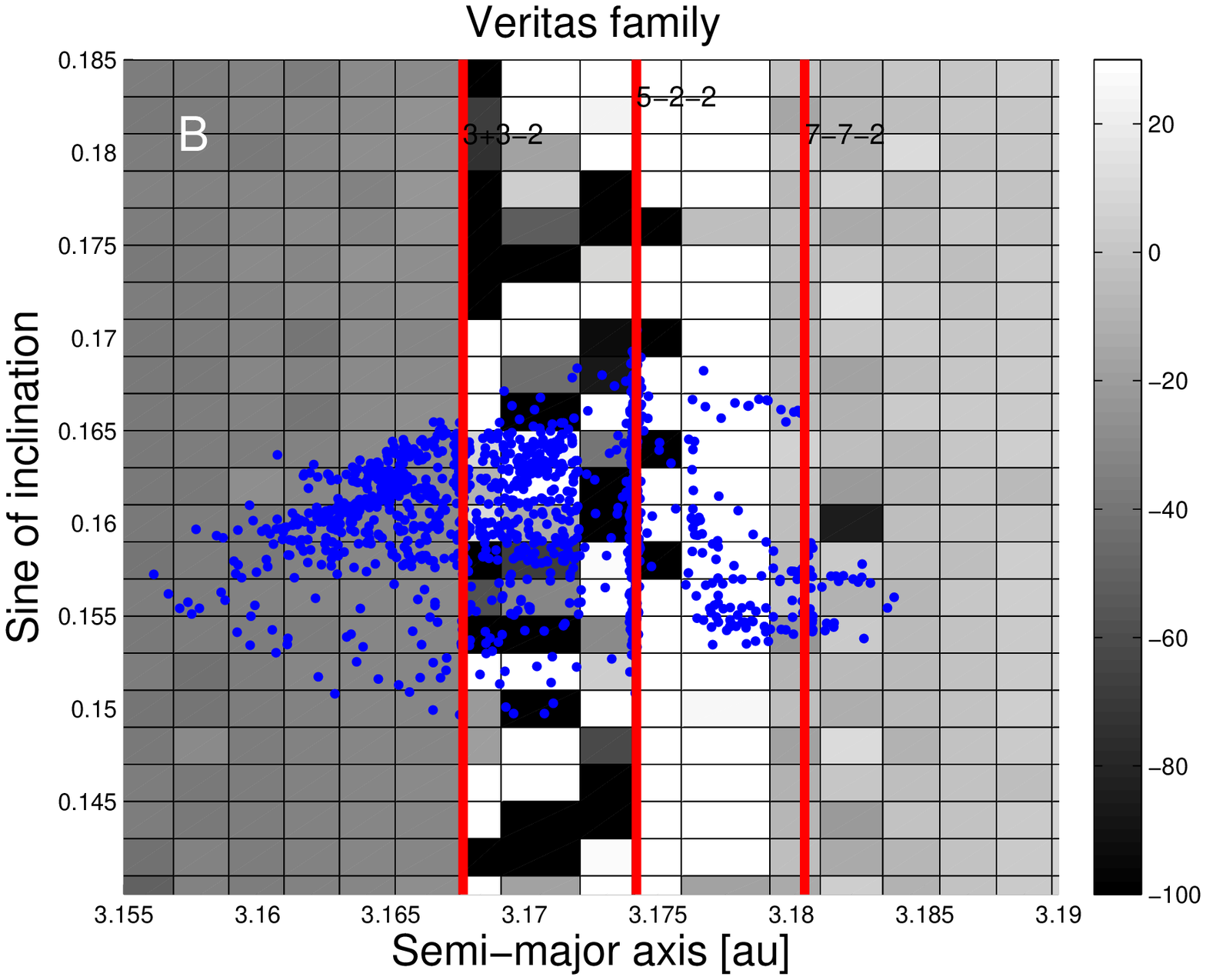}
  \end{minipage}
  \caption{An $(a,\sin{i})$ dynamical map of the region of the Veritas
family (Panel A).  Black dots identify the values of synthetic proper elements
of the integrated test particles.  Other symbols are the same as in 
Fig.~\ref{fig: veritas_back}.  Panel B displays a colour plot of values
of $\delta s/ \delta a$ obtained for the particles in the dynamical map
simulation.}
\label{fig: map_asg}
\end{figure*}

Following the approach of \citet{Nesvorny_2004}
we attempt to compute the drift rates caused by the Yarkovsky effect 
based on the values of $\Delta \Omega$, the difference
between the values of $\Omega$ between those of the 705
members of the Veritas family and those of (1086) Nata itself.
For this purpose, we first need to determine how the $g$ and $s$ 
proper frequencies depend on proper semi-major axis in the Veritas region.
Contrary to the case of the Karin cluster, we cannot use for the Veritas
family the convergence of the past longitudes of the nodes $\varpi$,
since the Veritas family is too close to the 2/1 mean-motion resonance
with Jupiter, and this causes the precession of the longitude
of the nodes to increase significantly. According to \citet{Nesvorny_2004},
values of $\Delta {\Omega}_{P,j}$ for a given j member at each given time
$t=-(\tau+\Delta t)$, with $\tau$ the estimated age of the Karin group,
is given by:

\begin{equation}
\Delta {\Omega}_{P,j}(t) ={\Omega}^{*}_{P,j}-{\Omega}^{*}_{P,1}
-\frac{1}{2}\frac{\partial s}{\partial {a}_P}(\delta a_{P,j}-
\Delta a_{P,1}) \tau -(s_j-s_1) \Delta t,
\label{eq: Delta_Om}
\end{equation}

where ${\Omega}^{*}_{P,j}-{\Omega}^{*}_{P,1}$ is the proper nodal
longitude difference caused by the ejection speeds $\Delta V$, and assumed
negligible hereafter (\citet{Nesvorny_2004} estimate that it is
of the order of $1^{\circ}$ for the Karin cluster, and should not be much
larger for the Veritas family).  $\frac{\partial s}{\partial {a}_P}$
define how frequencies change with $a_P$.  \citet{Nesvorny_2004}
used the analytic perturbation theory of \citet{Milani_1994} 
to estimate this rate equal to $-70.0$~arcsec/yr/au for all Karin
cluster members (with an uncertainty of 1\% caused by the spread in
proper $a_P$ of the cluster members).  The case of the Veritas
family is, however, much more challenging.

Fig.~\ref{fig: dsa} displays an $(a,s)$ projection of members of
the Karin cluster (panel A) and of the Veritas family (panel B).
While indeed most Karin members follows a single line of constant
$\delta s/ \delta a$ in the $(a,s)$ plane, the distribution of $(a,s)$
for the Veritas family is much more disperse.   The local three-body 
resonances play an important role in affecting the values of asteroid proper in
the region of the Veritas family, in a way not observed for the Karin
cluster.  To investigate the role of the local dynamics, we first 
obtained a dynamical map of synthetic proper elements in the
$(a,\sin{i})$ domain.

For this purpose, we created a grid of 1600 particles divided in 40 equally
spaced intervals in both osculating $a$ and $i$ and integrated them over 
12 Myr over the influence of the sun and the eight planets with 
$SWIFT\_MVSF$, the symplectic integrator
based on $SWIFT\_MVS$ from the {\em Swift} package of \citet{Levison_1994},
and modified by \citet{Broz_1999} to include online filtering of osculating
elements.  The initial osculating elements of the particles went
from 3.150 to 3.190~au in $a$ and from $7.0^{\circ}$ to $11.0^{\circ}$ in 
$i$.  The other orbital elements of the test particles were set equal to 
those of (1086) Nata at the modified Julian date of 57200.  The step
in osculating $a$, 0.001 au, was chosen small enough to allow for 
a significant resolution in the map, but large enough so that the computation
of $\delta s/ \delta a$ and $\delta g/ \delta a$ was precise enough,
considering the errors in proper frequencies and semi-major axis.
Synthetic proper elements were then obtained with the approach 
described in \citet{Carruba_2010}, based on the method of \citet{Knezevic_2003}.

Results are shown in Fig.~\ref{fig: map_asg}, panel A.   Black dots display
the location of the test particles synthetic proper elements.  
Not surprisingly, one can notice i) the important perturbing effect 
of the 5-2-2 resonance, and ii) the absence of important secular
resonances in this region, apart from the ${\nu}_{1H} = g-g_{Hygiea}$
located at $\simeq 3.170$~au (not shown in the figure for simplicity), 
that further contributes to the chaotic
dynamics between the 3+3-2 and 5-2-2 three-body resonances.  
Based on the values of proper frequencies
$g$ and $s$ obtained for this map, we then computed the values
of $\delta g/ \delta a$ and $\delta s /\delta a$ with this method: 
for each point in the line of 40 intervals in $a$ in the 
map, with the exception of the first and last, we computed the distance 
in proper $a$, $da$, and in proper frequencies $dg$ and $ds$
of the neighbor to the left with respect to the neighbor to the right. 
$\delta g/ \delta a$ and $\delta s /\delta a$
were then assumed equal to $dg/da$ and $ds/da$, respectively.  
Results are shown in Fig.~\ref{fig: map_asg}, panel B, display 
a color plot of our results for $\delta s /\delta a$ (results for
$\delta g/ \delta a$ are analogous and will not be shown, for 
simplicity).  For semi-major axis lower than those
of the center of the 3+3-2 resonance the behavior of $\delta s /\delta a$
is quite regular, slowly increasing with respect to $a$.  $\delta s /\delta a$
becomes much more erratic between the 3+3-2 and 5-2-2 resonances,
and only returns to a more regular behavior for values of semi-major
axis larger than 3.18~au, beyond the locations of the 5-2-2 and 
the 7-7-2 three-body resonances.  In view of the complex behavior
observed for $\delta s /\delta a$ and $\delta g/ \delta a$, 
we decided not to use an analytic approach to obtain family ages and 
drift rated based on these values.

\section{Appendix 2:  Yarkovsky drift speed values}
\label{sec: appendix_2}

We report in Table~\ref{table: Veritas_members} the first 10 
identified Veritas members, their absolute magnitude, proper 
$a, e, \sin{i}, g$, and $s$, Lyapunov exponent (multiplied by a 
factor $10^{6}$), and estimated mean
Yarkovsky drift speed, in au/Myr (no such value is available for
(1086) Nata itself, that because of its relative large size has very
limited Yarkovsky mobility). 

\newpage
\onecolumn
\begin{center}
\begin{longtable}{|c|c|c|c|c|c|c|c|c|}
  \caption{Proper elements and estimated drift rates of 274 Veritas family
    members.}
\label{table: Veritas_members}\\
\hline
Number & $H$ & $a_P$ & $e_P$ & $sin{i_P}$ & $g_P$ & $s_P$  & $LCE*10^{6}$& Drift speed \\
       &     & [au]  &       &           & [''/yr] & [''/yr] &  [yr$^{-1}$]&  [au/Myr] \\
\hline
\endfirsthead
\multicolumn{9}{|c|}%
{\tablename\ \thetable\ -- \textit{Continued from previous page}} \\
\hline
Number & $H$ & $a_P$ & $e_P$ & $sin{i_P}$ & $g_P$ & $s_P$  & $LCE*10^{6}$& Drift speed \\
       &     & [au]  &       &           & [''/yr] & [''/yr] &  [yr$^{-1}$]&  [au/Myr] \\
\hline
\endhead

\hline \multicolumn{9}{|r|}{{Continued on next page}} \\ \hline
\endfoot
\hline 
\endlastfoot
  1086 &   9.49 &   3.165357 &   0.061441 &   0.161970 & 127.944574 & -81.597922 &   1.45 &  0.000E+00 \\
  6374 &  11.95 &   3.165615 &   0.061238 &   0.161059 & 128.412635 & -81.663237 &   1.34 & -0.800E-04 \\
  7231 &  12.14 &   3.165802 &   0.061364 &   0.160469 & 128.750656 & -81.730850 &   0.33 & -0.220E-03 \\
  8624 &  13.55 &   3.164742 &   0.061074 &   0.162768 & 127.137085 & -81.472127 &   1.47 &  0.800E-03 \\
  9715 &  13.23 &   3.165485 &   0.060828 &   0.159609 & 128.625973 & -81.736346 &   1.87 & -0.740E-03 \\
  9860 &  13.07 &   3.164893 &   0.060683 &   0.161880 & 127.484952 & -81.514396 &   1.46 & -0.140E-03 \\
 11768 &  12.93 &   3.164016 &   0.060272 &   0.162164 & 126.564674 & -81.416291 &   1.50 &  0.300E-03 \\
 15066 &  12.23 &   3.165616 &   0.061469 &   0.162364 & 128.100351 & -81.579624 &   1.38 & -0.240E-03 \\
 15256 &  12.52 &   3.164626 &   0.060472 &   0.161926 & 127.209465 & -81.479886 &   1.52 &  0.820E-03 \\
 15732 &  12.38 &   3.164794 &   0.060991 &   0.162029 & 127.366566 & -81.526495 &   1.40 &  0.999E-04 \\
  17439 &  13.80 &   3.162809 &   0.059831 &   0.160005 & 125.957459 & -81.496475 &   1.70 &  0.200E-04 \\
 17701 &  12.44 &   3.165318 &   0.061097 &   0.162130 & 127.851683 & -81.550972 &   1.49 &  0.300E-02 \\
 18476 &  13.30 &   3.164524 &   0.060754 &   0.162374 & 127.014041 & -81.465351 &   1.44 & -0.800E-04 \\
 25828 &  13.02 &   3.164705 &   0.060953 &   0.161937 & 127.303391 & -81.526761 &   1.48 & -0.180E-03 \\
 28546 &  12.35 &   3.165455 &   0.061326 &   0.163226 & 127.722333 & -81.489204 &   1.54 &  0.680E-03 \\
 28908 &  13.10 &   3.164650 &   0.060736 &   0.162437 & 127.117559 & -81.464060 &   1.45 &  0.000E+00 \\
 31743 &  12.59 &   3.164265 &   0.060567 &   0.161117 & 127.071091 & -81.538253 &   1.50 &  0.500E-03 \\
 35041 &  14.81 &   3.164042 &   0.060291 &   0.159507 & 127.244037 & -81.632330 &   1.51 & -0.400E-03 \\
 37360 &  13.57 &   3.161214 &   0.058530 &   0.158279 & 124.892780 & -81.447510 &   1.47 &  0.300E-03 \\
 38419 &  13.84 &   3.163684 &   0.059980 &   0.160858 & 126.564888 & -81.480149 &   1.47 & -0.580E-03 \\
 42776 &  12.84 &   3.164625 &   0.060862 &   0.161532 & 127.324448 & -81.547888 &   1.48 &  0.120E-03 \\
 45727 &  13.41 &   3.163868 &   0.060452 &   0.160952 & 126.732618 & -81.523407 &   0.36 &  0.180E-03 \\
 46350 &  13.79 &   3.164370 &   0.060277 &   0.163091 & 126.669979 & -81.356635 &   1.45 &  0.160E-03 \\
 46412 &  14.05 &   3.163083 &   0.059002 &   0.160831 & 125.973686 & -81.367718 &   1.59 &  0.720E-03 \\
 47641 &  14.22 &   3.161093 &   0.058860 &   0.159794 & 124.435418 & -81.350636 &   1.09 &  0.000E+00 \\
 50182 &  13.04 &   3.164896 &   0.061141 &   0.162245 & 127.419207 & -81.527880 &   1.44 & -0.120E-03 \\
 51336 &  13.01 &   3.163305 &   0.059763 &   0.160663 & 126.253041 & -81.459604 &   1.35 & -0.660E-03 \\
 52551 &  13.62 &   3.164611 &   0.060742 &   0.162598 & 127.042241 & -81.449701 &   1.55 &  0.000E+00 \\
 54354 &  13.86 &   3.165539 &   0.061478 &   0.161905 & 128.141623 & -81.614246 &   1.49 &  0.460E-03 \\
 54592 &  13.89 &   3.164968 &   0.061088 &   0.161707 & 127.620458 & -81.569666 &   1.53 & -0.220E-03 \\
 57012 &  14.03 &   3.164837 &   0.060983 &   0.162208 & 127.365196 & -81.513115 &   1.33 &  0.800E-04 \\
 57753 &  14.44 &   3.164231 &   0.060510 &   0.162583 & 126.675971 & -81.413634 &   1.47 &  0.440E-03 \\
 57950 &  14.05 &   3.161606 &   0.059459 &   0.158770 & 125.160602 & -81.508210 &   0.18 &  0.520E-03 \\
 60256 &  14.01 &   3.164504 &   0.060246 &   0.161217 & 127.260845 & -81.511176 &   1.48 & -0.120E-03 \\
 60329 &  13.96 &   3.163761 &   0.060284 &   0.160477 & 126.744086 & -81.541785 &   1.68 &  0.140E-03 \\
 60605 &  13.59 &   3.164496 &   0.060418 &   0.163034 & 126.810078 & -81.379503 &   1.54 &  0.600E-03 \\
 62467 &  13.92 &   3.162568 &   0.059634 &   0.159219 & 125.922575 & -81.531169 &   0.28 & -0.340E-03 \\
 72563 &  13.78 &   3.164477 &   0.060664 &   0.160876 & 127.337461 & -81.576150 &   1.52 &  0.999E-04 \\
 72819 &  14.13 &   3.164171 &   0.060300 &   0.159489 & 127.371670 & -81.640429 &   1.60 &  0.860E-03 \\
 78726 &  14.55 &   3.164401 &   0.060356 &   0.163143 & 126.690025 & -81.360469 &   1.49 &  0.540E-03 \\
 81935 &  13.61 &   3.164941 &   0.061000 &   0.162855 & 127.306185 & -81.467112 &   1.49 & -0.440E-03 \\
 82961 &  13.97 &   3.163596 &   0.059972 &   0.160924 & 126.467789 & -81.470525 &   1.56 &  0.220E-03 \\
 83253 &  14.21 &   3.164422 &   0.060193 &   0.157813 & 128.014094 & -81.774563 &   1.74 & -0.300E-03 \\
 83802 &  14.38 &   3.161932 &   0.059400 &   0.160545 & 125.018547 & -81.375745 &   1.28 &  0.580E-03 \\
 86285 &  13.71 &   3.165185 &   0.060962 &   0.161573 & 127.858357 & -81.577812 &   1.50 & -0.340E-03 \\
 87898 &  14.37 &   3.164912 &   0.060382 &   0.162060 & 127.448558 & -81.473145 &   0.04 & -0.460E-03 \\
 90252 &  14.64 &   3.162968 &   0.059721 &   0.159632 & 126.192638 & -81.523771 &   1.25 & -0.400E-03 \\
 91850 &  14.51 &   3.160705 &   0.058510 &   0.159686 & 124.110854 & -81.311467 &   1.63 &  0.620E-03 \\
 92131 &  14.20 &   3.163030 &   0.059763 &   0.158249 & 126.586142 & -81.640068 &   1.39 & -0.779E-03 \\
 94167 &  14.48 &   3.162193 &   0.059465 &   0.160524 & 125.260672 & -81.395188 &   1.56 & -0.580E-03 \\
 94285 &  14.13 &   3.165299 &   0.061409 &   0.162704 & 127.707209 & -81.532720 &   1.49 &  0.122E-02 \\
 95505 &  14.72 &   3.160562 &   0.058394 &   0.154249 & 125.270716 & -81.720492 &   1.27 &  0.600E-03 \\
 96549 &  14.34 &   3.163247 &   0.059840 &   0.158451 & 126.740241 & -81.640343 &   1.42 & -0.180E-03 \\
 97654 &  13.94 &   3.163692 &   0.060497 &   0.161069 & 126.544221 & -81.510822 &   1.74 &  0.740E-03 \\
105854 &  13.98 &   3.164814 &   0.061434 &   0.161874 & 127.447695 & -81.581781 &   1.47 &  0.420E-03 \\
106551 &  14.44 &   3.157830 &   0.055532 &   0.155412 & 122.585895 & -81.262210 &   1.36 & -0.220E-03 \\
110595 &  14.38 &   3.162904 &   0.057505 &   0.150796 & 128.140527 & -82.010671 &   1.44 & -0.940E-03 \\
111713 &  14.74 &   3.164573 &   0.060242 &   0.162772 & 126.940836 & -81.387432 &   1.44 & -0.180E-03 \\
112107 &  14.40 &   3.163064 &   0.059862 &   0.160615 & 126.047631 & -81.462291 &   1.40 &  0.200E-03 \\
113283 &  14.07 &   3.164308 &   0.060750 &   0.162749 & 126.719298 & -81.425637 &   1.20 &  0.000E+00 \\
115211 &  14.00 &   3.164560 &   0.060443 &   0.162713 & 126.952903 & -81.410589 &   1.47 &  0.200E-04 \\
127336 &  14.21 &   3.161449 &   0.059159 &   0.160294 & 124.640439 & -81.352844 &   0.51 &  0.280E-03 \\
129117 &  14.86 &   3.157537 &   0.055855 &   0.155113 & 122.429214 & -81.298008 &   0.43 &  0.200E-03 \\
130355 &  14.91 &   3.165214 &   0.060437 &   0.158157 & 128.702083 & -81.803724 &   1.41 & -0.620E-03 \\
130360 &  14.66 &   3.164541 &   0.060989 &   0.162446 & 127.024392 & -81.482036 &   0.50 &  0.400E-04 \\
132750 &  13.73 &   3.165298 &   0.061292 &   0.161462 & 128.009493 & -81.622310 &   1.50 &  0.000E+00 \\
132849 &  14.25 &   3.164530 &   0.060361 &   0.159409 & 127.734302 & -81.667893 &   1.49 & -0.200E-04 \\
132861 &  14.29 &   3.164179 &   0.060180 &   0.161037 & 126.994440 & -81.505254 &   1.38 &  0.200E-04 \\
135404 &  14.42 &   3.164697 &   0.060773 &   0.161383 & 127.427027 & -81.554788 &   1.19 &  0.000E+00 \\
135416 &  14.49 &   3.160482 &   0.058240 &   0.159432 & 123.966463 & -81.298099 &   1.55 & -0.520E-03 \\
138569 &  14.80 &   3.163351 &   0.059506 &   0.160743 & 126.264671 & -81.432080 &   1.52 &  0.220E-03 \\
138747 &  14.51 &   3.165347 &   0.060746 &   0.159569 & 128.499718 & -81.725999 &   1.44 &  0.240E-03 \\
138758 &  14.38 &   3.165422 &   0.061351 &   0.161845 & 128.036724 & -81.601904 &   1.41 &  0.108E-02 \\
141189 &  14.62 &   3.163511 &   0.060015 &   0.160236 & 126.557860 & -81.525640 &   1.79 &  0.420E-03 \\
144405 &  14.62 &   3.165479 &   0.061113 &   0.162716 & 127.865280 & -81.511759 &   0.00 & -0.580E-03 \\
145894 &  14.66 &   3.165646 &   0.061088 &   0.161916 & 128.226823 & -81.581238 &   1.51 &  0.700E-03 \\
147948 &  14.90 &   3.164483 &   0.060393 &   0.162742 & 126.868661 & -81.399981 &   1.48 & -0.120E-03 \\
149141 &  14.78 &   3.161778 &   0.059075 &   0.160083 & 124.979202 & -81.376913 &   1.25 &  0.800E-04 \\
150320 &  14.01 &   3.164077 &   0.060115 &   0.161041 & 126.894299 & -81.494536 &   1.43 & -0.779E-03 \\
153169 &  14.61 &   3.163763 &   0.060351 &   0.162197 & 126.326428 & -81.410094 &   1.58 & -0.400E-04 \\
155231 &  15.00 &   3.163095 &   0.059924 &   0.159959 & 126.237566 & -81.521071 &   0.87 & -0.160E-03 \\
155463 &  14.70 &   3.164984 &   0.061019 &   0.162578 & 127.416392 & -81.492734 &   1.47 &  0.600E-04 \\
157463 &  14.52 &   3.163610 &   0.059855 &   0.160753 & 126.516740 & -81.474283 &   1.48 &  0.999E-04 \\
162156 &  14.84 &   3.165859 &   0.061219 &   0.164015 & 127.915770 & -81.431502 &   1.45 &  0.150E-02 \\
168041 &  15.17 &   3.164292 &   0.060234 &   0.162945 & 126.630872 & -81.361212 &   1.49 & -0.520E-03 \\
169282 &  14.50 &   3.164507 &   0.060058 &   0.162416 & 126.957856 & -81.396788 &   1.49 &  0.600E-04 \\
169314 &  14.81 &   3.162605 &   0.059784 &   0.160395 & 125.677826 & -81.452821 &   1.14 &  0.116E-02 \\
169398 &  14.75 &   3.164175 &   0.060304 &   0.159836 & 127.290531 & -81.613062 &   1.51 &  0.999E-04 \\
176576 &  14.63 &   3.164367 &   0.060320 &   0.160172 & 127.389730 & -81.595561 &   1.50 & -0.560E-03 \\
178745 &  14.59 &   3.162774 &   0.059407 &   0.160669 & 125.747721 & -81.403791 &   1.28 &  0.000E+00 \\
179448 &  14.66 &   3.164581 &   0.060632 &   0.162481 & 127.036884 & -81.447747 &   1.15 &  0.400E-04 \\
181265 &  15.69 &   3.162109 &   0.057693 &   0.163171 & 124.465833 & -81.022891 &   1.47 &  0.620E-03 \\
181598 &  15.19 &   3.163722 &   0.058446 &   0.152527 & 128.536483 & -81.997301 &   1.53 &  0.500E-03 \\
186634 &  15.01 &   3.164985 &   0.060455 &   0.161774 & 127.590964 & -81.506008 &   1.49 & -0.800E-04 \\
188944 &  14.87 &   3.163005 &   0.059501 &   0.162372 & 125.545823 & -81.285939 &   1.79 & -0.999E-04 \\
191443 &  14.43 &   3.162569 &   0.059473 &   0.160662 & 125.565330 & -81.401444 &   0.26 &  0.680E-03 \\
192460 &  15.03 &   3.163392 &   0.059904 &   0.160924 & 126.274037 & -81.455694 &   1.51 &  0.999E-04 \\
192604 &  15.48 &   3.165851 &   0.060891 &   0.159678 & 128.972717 & -81.752176 &   1.52 &  0.120E-02 \\
196134 &  14.79 &   3.165773 &   0.060646 &   0.157670 & 129.377797 & -81.886030 &   1.50 & -0.800E-04 \\
197076 &  15.21 &   3.164561 &   0.060107 &   0.163156 & 126.826630 & -81.343886 &   1.32 & -0.400E-04 \\
198745 &  14.86 &   3.165625 &   0.060803 &   0.158543 & 129.023528 & -81.824357 &   1.28 & -0.220E-03 \\
200365 &  14.88 &   3.161320 &   0.058877 &   0.159895 & 124.609717 & -81.353918 &   1.24 &  0.400E-04 \\
202319 &  15.11 &   3.163660 &   0.059915 &   0.162666 & 126.092772 & -81.327007 &   0.95 &  0.560E-03 \\
207576 &  15.54 &   3.164195 &   0.060311 &   0.159170 & 127.471084 & -81.667825 &   1.41 & -0.140E-03 \\
208453 &  14.87 &   3.165235 &   0.061341 &   0.160429 & 128.203393 & -81.707337 &   1.38 &  0.380E-03 \\
209410 &  15.23 &   3.158645 &   0.056316 &   0.156275 & 123.082688 & -81.297274 &   0.48 &  0.900E-03 \\
210067 &  15.40 &   3.164961 &   0.060973 &   0.160719 & 127.850600 & -81.637836 &   1.43 &  0.258E-02 \\
210171 &  15.32 &   3.161992 &   0.059545 &   0.158566 & 125.554539 & -81.549477 &   1.48 &  0.620E-03 \\
210431 &  15.13 &   3.165053 &   0.060528 &   0.158316 & 128.509419 & -81.792582 &   1.49 &  0.300E-03 \\
214439 &  15.84 &   3.164329 &   0.060334 &   0.161223 & 127.094467 & -81.510907 &   1.62 & -0.380E-03 \\
214560 &  15.59 &   3.164763 &   0.061019 &   0.162386 & 127.250341 & -81.499325 &   1.51 & -0.180E-03 \\
215011 &  15.77 &   3.165512 &   0.059723 &   0.164306 & 127.435501 & -81.255152 &   1.79 & -0.400E-04 \\
216775 &  14.85 &   3.165294 &   0.061470 &   0.162832 & 127.670473 & -81.527715 &   1.58 &  0.274E-02 \\
217140 &  15.24 &   3.165071 &   0.060921 &   0.162636 & 127.478097 & -81.482940 &   1.65 &  0.460E-03 \\
217391 &  15.13 &   3.164550 &   0.060301 &   0.159361 & 127.759919 & -81.666721 &   0.00 & -0.360E-03 \\
220777 &  14.44 &   3.164048 &   0.059943 &   0.161029 & 126.859958 & -81.478476 &   1.41 &  0.000E+00 \\
223398 &  15.91 &   3.160882 &   0.058613 &   0.159697 & 124.262112 & -81.327265 &   0.83 &  0.400E-04 \\
223862 &  15.22 &   3.161153 &   0.058558 &   0.158288 & 124.836266 & -81.445759 &   1.29 &  0.000E+00 \\
225150 &  15.50 &   3.164971 &   0.060411 &   0.161581 & 127.621764 & -81.516916 &   1.48 & -0.800E-03 \\
226517 &  14.81 &   3.165446 &   0.060499 &   0.163492 & 127.608311 & -81.389947 &   1.46 & -0.340E-03 \\
232975 &  15.04 &   3.165695 &   0.061124 &   0.160842 & 128.539577 & -81.673463 &   0.38 &  0.180E-03 \\
233908 &  14.84 &   3.165270 &   0.060492 &   0.163719 & 127.380312 & -81.362877 &   1.48 &  0.108E-02 \\
234440 &  15.56 &   3.163900 &   0.059546 &   0.160984 & 126.715039 & -81.439944 &   1.34 & -0.500E-03 \\
236668 &  16.14 &   3.163575 &   0.059868 &   0.162278 & 126.106596 & -81.350946 &   1.49 & -0.580E-03 \\
236811 &  15.51 &   3.158787 &   0.057676 &   0.155852 & 123.351490 & -81.452657 &   1.47 &  0.000E+00 \\
236970 &  14.72 &   3.163826 &   0.060369 &   0.160669 & 126.757684 & -81.536676 &   1.38 &  0.460E-03 \\
237363 &  15.44 &   3.160700 &   0.057110 &   0.153021 & 125.623586 & -81.708655 &   1.49 &  0.400E-04 \\
238392 &  15.72 &   3.163708 &   0.060423 &   0.161094 & 126.545628 & -81.502554 &   1.49 &  0.999E-04 \\
238402 &  15.74 &   3.164935 &   0.061329 &   0.160998 & 127.771335 & -81.647583 &   1.49 &  0.360E-03 \\
239217 &  15.44 &   3.164584 &   0.059879 &   0.162759 & 126.935285 & -81.355890 &   1.74 & -0.200E-03 \\
240266 &  15.07 &   3.165718 &   0.061261 &   0.163636 & 127.870964 & -81.460219 &   1.72 &  0.460E-03 \\
240533 &  15.26 &   3.165634 &   0.061105 &   0.159050 & 128.921164 & -81.812696 &   1.34 & -0.800E-04 \\
241998 &  15.14 &   3.162766 &   0.059713 &   0.157700 & 126.469989 & -81.666927 &   1.55 & -0.600E-04 \\
242017 &  15.50 &   3.161123 &   0.057848 &   0.153468 & 125.920053 & -81.757651 &   1.37 &  0.800E-04 \\
242288 &  14.98 &   3.164966 &   0.060766 &   0.158795 & 128.317605 & -81.772948 &   1.48 &  0.400E-03 \\
242675 &  15.36 &   3.162021 &   0.059653 &   0.159677 & 125.316910 & -81.471837 &   0.75 &  0.126E-02 \\
242959 &  15.09 &   3.165335 &   0.061364 &   0.163139 & 127.626949 & -81.494192 &   1.52 &  0.000E+00 \\
244159 &  15.30 &   3.163505 &   0.059877 &   0.159672 & 126.680950 & -81.557847 &   1.61 & -0.940E-03 \\
245064 &  15.88 &   3.165459 &   0.066143 &   0.162188 & 128.211747 & -82.045206 &   1.24 & -0.140E-03 \\
245488 &  15.55 &   3.164897 &   0.060454 &   0.161478 & 127.577404 & -81.525969 &   1.58 & -0.700E-03 \\
245494 &  14.87 &   3.164868 &   0.059949 &   0.163625 & 126.994682 & -81.304405 &   1.50 &  0.280E-03 \\
246756 &  16.21 &   3.163959 &   0.058865 &   0.153337 & 128.585278 & -81.983080 &   1.43 & -0.140E-03 \\
248121 &  15.31 &   3.164166 &   0.059996 &   0.161251 & 126.918623 & -81.471037 &   1.56 &  0.360E-03 \\
249273 &  16.00 &   3.163570 &   0.058890 &   0.163030 & 125.874804 & -81.201645 &   1.19 & -0.112E-02 \\
249492 &  15.50 &   3.162780 &   0.059558 &   0.158085 & 126.383466 & -81.622981 &   1.69 &  0.400E-04 \\
250082 &  15.28 &   3.163844 &   0.060025 &   0.163399 & 126.086687 & -81.285654 &   1.54 &  0.720E-03 \\
251802 &  16.00 &   3.165060 &   0.060615 &   0.158457 & 128.483955 & -81.789748 &   1.48 &  0.999E-04 \\
251842 &  15.89 &   3.164093 &   0.060188 &   0.157388 & 127.800984 & -81.793108 &   0.38 &  0.000E+00 \\
252834 &  15.41 &   3.161034 &   0.058779 &   0.159778 & 124.382255 & -81.342040 &   0.00 &  0.999E-04 \\
254391 &  15.52 &   3.162598 &   0.059654 &   0.160165 & 125.719198 & -81.458722 &   1.89 &  0.400E-04 \\
255243 &  14.98 &   3.163279 &   0.060096 &   0.160718 & 126.226480 & -81.484260 &   1.49 & -0.360E-03 \\
255332 &  14.97 &   3.163638 &   0.060332 &   0.159539 & 126.857059 & -81.616191 &   0.28 &  0.800E-04 \\
257098 &  15.21 &   3.164292 &   0.060765 &   0.161463 & 127.018818 & -81.529931 &   1.45 &  0.779E-03 \\
258829 &  15.21 &   3.163789 &   0.060245 &   0.158904 & 127.149357 & -81.665102 &   1.50 &  0.820E-03 \\
261446 &  15.94 &   3.165503 &   0.060245 &   0.161783 & 128.078475 & -81.507648 &   1.73 & -0.132E-02 \\
261464 &  15.70 &   3.162839 &   0.060209 &   0.159711 & 126.072233 & -81.556055 &   1.53 &  0.000E+00 \\
261618 &  15.39 &   3.162828 &   0.059794 &   0.158380 & 126.367283 & -81.623442 &   1.49 & -0.540E-03 \\
261709 &  14.83 &   3.161433 &   0.059111 &   0.160159 & 124.653800 & -81.359114 &   1.42 &  0.440E-03 \\
263510 &  15.80 &   3.162995 &   0.059260 &   0.157257 & 126.767162 & -81.670741 &   1.49 & -0.600E-03 \\
264821 &  14.91 &   3.164648 &   0.060329 &   0.161236 & 127.393036 & -81.522619 &   1.47 & -0.220E-03 \\
265059 &  15.79 &   3.165989 &   0.060934 &   0.163733 & 128.099557 & -81.433257 &   0.43 & -0.720E-03 \\
265082 &  14.95 &   3.161647 &   0.058775 &   0.159106 & 125.084788 & -81.422318 &   0.00 &  0.999E-04 \\
265109 &  15.04 &   3.165357 &   0.061343 &   0.162340 & 127.847634 & -81.558692 &   1.37 &  0.112E-02 \\
265446 &  14.99 &   3.163480 &   0.060108 &   0.160053 & 126.575615 & -81.547300 &   1.39 & -0.560E-03 \\
268202 &  15.69 &   3.158716 &   0.057002 &   0.159210 & 122.474700 & -81.129063 &   1.19 &  0.000E+00 \\
268204 &  16.16 &   3.164260 &   0.060228 &   0.158899 & 127.593441 & -81.684279 &   1.53 &  0.000E+00 \\
269924 &  15.58 &   3.162917 &   0.059756 &   0.160118 & 126.024947 & -81.485324 &   0.31 &  0.640E-03 \\
270913 &  15.56 &   3.162643 &   0.059205 &   0.162654 & 125.132029 & -81.220354 &   1.47 & -0.520E-03 \\
271223 &  16.48 &   3.162183 &   0.059181 &   0.155955 & 126.335455 & -81.730302 &   0.00 &  0.108E-02 \\
272045 &  15.41 &   3.164130 &   0.059526 &   0.161240 & 126.867470 & -81.427413 &   1.53 &  0.400E-04 \\
272073 &  15.55 &   3.165127 &   0.059809 &   0.164110 & 127.114848 & -81.262976 &   1.42 &  0.360E-03 \\
274064 &  15.43 &   3.164794 &   0.060375 &   0.161382 & 127.499445 & -81.522054 &   1.48 & -0.580E-03 \\
274752 &  15.48 &   3.165504 &   0.061692 &   0.161935 & 128.107137 & -81.630518 &   1.06 &  0.120E-03 \\
275175 &  15.60 &   3.161495 &   0.059023 &   0.159929 & 124.761600 & -81.372206 &   0.79 & -0.280E-03 \\
276667 &  15.55 &   3.163017 &   0.058711 &   0.161325 & 125.778854 & -81.299834 &   1.77 & -0.480E-03 \\
279094 &  14.73 &   3.164100 &   0.059880 &   0.162286 & 126.595862 & -81.373552 &   1.30 &  0.999E-04 \\
279368 &  15.55 &   3.165825 &   0.059627 &   0.154176 & 130.227072 & -82.067778 &   1.55 & -0.120E-03 \\
279436 &  15.19 &   3.162593 &   0.059216 &   0.160613 & 125.587276 & -81.383384 &   0.94 &  0.880E-03 \\
279499 &  16.41 &   3.165116 &   0.060281 &   0.156714 & 128.947909 & -81.899027 &   1.63 &  0.420E-03 \\
280167 &  15.30 &   3.163826 &   0.060624 &   0.161277 & 126.619650 & -81.511388 &   1.53 &  0.600E-03 \\
281319 &  15.36 &   3.162761 &   0.059511 &   0.159742 & 125.964370 & -81.486635 &   1.68 &  0.160E-03 \\
282384 &  14.94 &   3.163622 &   0.059753 &   0.157618 & 127.282963 & -81.714616 &   1.36 &  0.600E-03 \\
283206 &  15.54 &   3.163475 &   0.060202 &   0.160776 & 126.397973 & -81.498000 &   1.53 &  0.200E-03 \\
284839 &  15.29 &   3.163418 &   0.059624 &   0.160720 & 126.337973 & -81.447818 &   1.07 & -0.400E-03 \\
284916 &  15.65 &   3.163443 &   0.060171 &   0.160625 & 126.403503 & -81.505721 &   1.56 &  0.118E-02 \\
285109 &  14.87 &   3.164236 &   0.060586 &   0.163234 & 126.518840 & -81.367510 &   1.81 &  0.128E-02 \\
285135 &  15.64 &   3.163752 &   0.059850 &   0.160395 & 126.733851 & -81.508312 &   1.29 & -0.740E-03 \\
290614 &  16.16 &   3.164968 &   0.060752 &   0.161216 & 127.724092 & -81.577593 &   1.55 & -0.200E-03 \\
290828 &  16.49 &   3.164837 &   0.060467 &   0.163868 & 126.925442 & -81.330094 &   1.54 &  0.142E-02 \\
292125 &  15.93 &   3.164360 &   0.059792 &   0.162810 & 126.706765 & -81.334231 &   1.49 &  0.580E-03 \\
292308 &  15.53 &   3.165903 &   0.061318 &   0.161773 & 128.521575 & -81.625243 &   1.41 & -0.460E-03 \\
294550 &  15.55 &   3.162395 &   0.059787 &   0.157771 & 126.118251 & -81.651333 &   1.55 &  0.184E-02 \\
294595 &  15.49 &   3.164692 &   0.060991 &   0.162565 & 127.135632 & -81.478886 &   1.75 &  0.340E-03 \\
294648 &  15.60 &   3.165151 &   0.060691 &   0.162596 & 127.555129 & -81.468502 &   1.52 & -0.142E-02 \\
294679 &  15.88 &   3.162807 &   0.059461 &   0.158989 & 126.186503 & -81.543748 &   1.58 & -0.300E-03 \\
295434 &  16.32 &   3.165106 &   0.060814 &   0.162282 & 127.595498 & -81.503171 &   1.52 & -0.480E-03 \\
296630 &  15.65 &   3.163898 &   0.059894 &   0.161000 & 126.724389 & -81.470412 &   1.23 &  0.400E-04 \\
296907 &  15.15 &   3.162259 &   0.059159 &   0.162486 & 124.825102 & -81.212471 &   1.51 &  0.260E-03 \\
296992 &  15.97 &   3.163334 &   0.059873 &   0.162633 & 125.796713 & -81.311911 &   1.25 &  0.800E-03 \\
299255 &  16.00 &   3.163117 &   0.059725 &   0.160711 & 126.063377 & -81.443809 &   1.51 &  0.640E-03 \\
299298 &  15.29 &   3.161891 &   0.059605 &   0.157875 & 125.632031 & -81.604579 &   1.48 &  0.880E-03 \\
299320 &  16.04 &   3.165256 &   0.061116 &   0.159802 & 128.368339 & -81.737822 &   1.62 &  0.700E-03 \\
300116 &  15.63 &   3.162525 &   0.058826 &   0.158902 & 125.923633 & -81.481553 &   0.24 & -0.580E-03 \\
300181 &  16.09 &   3.161305 &   0.057411 &   0.160545 & 124.380538 & -81.173678 &   0.69 &  0.400E-04 \\
300195 &  15.00 &   3.161911 &   0.058860 &   0.161893 & 124.647411 & -81.218865 &   0.00 &  0.920E-03 \\
300866 &  15.06 &   3.163028 &   0.059724 &   0.158290 & 126.568884 & -81.632876 &   0.00 & -0.640E-03 \\
300909 &  15.36 &   3.161764 &   0.059116 &   0.162675 & 124.334935 & -81.172024 &   1.49 &  0.779E-03 \\
303795 &  16.19 &   3.159267 &   0.057610 &   0.159627 & 122.857829 & -81.172798 &   2.00 & -0.640E-03 \\
306092 &  16.06 &   3.165384 &   0.060745 &   0.162211 & 127.878895 & -81.514098 &   1.56 & -0.102E-02 \\
306983 &  15.30 &   3.165449 &   0.060594 &   0.157721 & 129.042951 & -81.863350 &   1.14 &  0.340E-03 \\
308390 &  15.59 &   3.158472 &   0.057217 &   0.159333 & 122.252852 & -81.126933 &   0.00 & -0.440E-03 \\
308419 &  15.27 &   3.164179 &   0.060792 &   0.162802 & 126.582147 & -81.418880 &   1.45 &  0.440E-03 \\
308493 &  16.16 &   3.165524 &   0.061208 &   0.160170 & 128.542529 & -81.727987 &   1.50 &  0.204E-02 \\
315842 &  15.68 &   3.157401 &   0.055411 &   0.155736 & 122.155088 & -81.206610 &   1.57 &  0.300E-03 \\
316525 &  15.49 &   3.164097 &   0.060834 &   0.162869 & 126.490692 & -81.414344 &   1.73 &  0.800E-03 \\
318349 &  15.66 &   3.164527 &   0.064169 &   0.151597 & 129.786494 & -82.649432 &   1.40 & -0.800E-04 \\
319165 &  15.66 &   3.165422 &   0.061074 &   0.162303 & 127.907391 & -81.538572 &   1.42 & -0.440E-03 \\
320061 &  15.74 &   3.165483 &   0.061365 &   0.164066 & 127.539477 & -81.425153 &   0.35 &  0.600E-03 \\
322203 &  15.83 &   3.165699 &   0.058617 &   0.150899 & 130.837567 & -82.222504 &   1.92 &  0.240E-02 \\
323274 &  15.74 &   3.163064 &   0.059469 &   0.161069 & 125.915541 & -81.389830 &   1.47 &  0.800E-03 \\
324281 &  15.99 &   3.164318 &   0.060670 &   0.162564 & 126.767537 & -81.433510 &   1.66 &  0.300E-03 \\
324754 &  16.09 &   3.164700 &   0.060102 &   0.161305 & 127.416795 & -81.499257 &   1.30 & -0.460E-03 \\
325369 &  15.45 &   3.165509 &   0.060856 &   0.158733 & 128.865128 & -81.809477 &   1.52 & -0.600E-04 \\
326258 &  15.22 &   3.165193 &   0.060897 &   0.161913 & 127.774356 & -81.544110 &   1.49 & -0.600E-03 \\
328045 &  16.19 &   3.165402 &   0.061389 &   0.163612 & 127.575281 & -81.461305 &   1.42 &  0.246E-02 \\
333826 &  15.83 &   3.163153 &   0.059920 &   0.159504 & 126.398803 & -81.559812 &   1.41 &  0.800E-04 \\
336481 &  15.88 &   3.164781 &   0.060901 &   0.162270 & 127.290717 & -81.498596 &   1.42 & -0.146E-02 \\
340685 &  15.38 &   3.164659 &   0.063509 &   0.161752 & 127.421121 & -81.782518 &   1.69 & -0.620E-03 \\
341786 &  15.31 &   3.165139 &   0.061075 &   0.163770 & 127.267399 & -81.407801 &   0.61 &  0.180E-02 \\
343046 &  15.65 &   3.164041 &   0.060351 &   0.161012 & 126.875341 & -81.516932 &   1.50 & -0.140E-03 \\
343049 &  15.59 &   3.162848 &   0.059614 &   0.158463 & 126.358237 & -81.601377 &   1.54 & -0.700E-03 \\
343575 &  16.32 &   3.165058 &   0.059381 &   0.164406 & 126.955566 & -81.196880 &   1.52 & -0.500E-03 \\
343595 &  15.44 &   3.157092 &   0.055573 &   0.155409 & 121.988603 & -81.231548 &   1.54 &  0.600E-04 \\
344374 &  15.36 &   3.163667 &   0.059935 &   0.162254 & 126.201344 & -81.362347 &   1.45 &  0.112E-02 \\
345648 &  15.98 &   3.162758 &   0.059920 &   0.162310 & 125.351615 & -81.317593 &   1.14 &  0.660E-03 \\
346796 &  16.33 &   3.165287 &   0.058259 &   0.151276 & 130.326147 & -82.143604 &   1.19 & -0.660E-03 \\
346835 &  15.88 &   3.164113 &   0.060119 &   0.158943 & 127.439879 & -81.664048 &   1.57 &  0.200E-04 \\
349308 &  15.61 &   3.164205 &   0.059705 &   0.160838 & 127.045273 & -81.479340 &   1.71 & -0.800E-03 \\
350123 &  16.57 &   3.165857 &   0.061487 &   0.159867 & 128.958052 & -81.793218 &   1.44 & -0.260E-03 \\
350804 &  16.29 &   3.163354 &   0.058801 &   0.153874 & 127.889029 & -81.909963 &   1.45 & -0.160E-03 \\
351936 &  16.35 &   3.164296 &   0.060155 &   0.161315 & 127.032968 & -81.485795 &   1.45 & -0.380E-03 \\
351967 &  15.50 &   3.160332 &   0.057891 &   0.159704 & 123.756953 & -81.239161 &   0.57 &  0.300E-03 \\
352544 &  15.93 &   3.162226 &   0.058204 &   0.160637 & 125.209410 & -81.275902 &   0.36 & -0.560E-03 \\
354702 &  16.21 &   3.161734 &   0.059102 &   0.162556 & 124.337648 & -81.179040 &   1.71 &  0.300E-03 \\
355679 &  15.34 &   3.164887 &   0.060638 &   0.163388 & 127.101979 & -81.387406 &   1.45 &  0.600E-03 \\
356365 &  15.79 &   3.163559 &   0.060594 &   0.160053 & 126.670234 & -81.595212 &   1.79 &  0.560E-03 \\
360117 &  15.60 &   3.163673 &   0.059720 &   0.160923 & 126.526762 & -81.450891 &   1.44 &  0.440E-03 \\
361491 &  15.66 &   3.159413 &   0.058016 &   0.157061 & 123.606657 & -81.416280 &   0.13 &  0.300E-03 \\
362477 &  15.88 &   3.165544 &   0.060626 &   0.157818 & 129.114604 & -81.862845 &   1.55 &  0.480E-03 \\
362566 &  15.51 &   3.162657 &   0.059276 &   0.160484 & 125.679785 & -81.401422 &   1.46 &  0.140E-03 \\
363484 &  15.44 &   3.164353 &   0.059763 &   0.162674 & 126.734586 & -81.342618 &   1.48 & -0.520E-03 \\
363869 &  15.88 &   3.162682 &   0.059973 &   0.162599 & 125.214211 & -81.295715 &   1.23 &  0.204E-02 \\
364562 &  16.17 &   3.165703 &   0.060025 &   0.163580 & 127.815399 & -81.349777 &   0.19 & -0.136E-02 \\
365379 &  15.85 &   3.163508 &   0.058330 &   0.154015 & 127.978952 & -81.862631 &   1.25 & -0.880E-03 \\
365514 &  15.73 &   3.161317 &   0.058618 &   0.159061 & 124.798319 & -81.397986 &   0.58 &  0.116E-02 \\
365809 &  15.82 &   3.163723 &   0.060165 &   0.161034 & 126.564824 & -81.484545 &   1.27 &  0.900E-03 \\
366523 &  16.16 &   3.159244 &   0.056049 &   0.154123 & 124.076467 & -81.467856 &   0.00 & -0.520E-03 \\
366544 &  16.45 &   3.162277 &   0.057536 &   0.151725 & 127.349253 & -81.914754 &   1.43 & -0.200E-03 \\
366804 &  16.19 &   3.165315 &   0.061056 &   0.160385 & 128.279357 & -81.687629 &   1.50 & -0.420E-03 \\
368082 &  15.90 &   3.163981 &   0.059517 &   0.161034 & 126.777784 & -81.436465 &   0.37 &  0.120E-03 \\
369180 &  16.24 &   3.164120 &   0.060602 &   0.161870 & 126.750064 & -81.474264 &   1.44 &  0.500E-03 \\
370407 &  16.14 &   3.161200 &   0.059732 &   0.159108 & 124.729105 & -81.487202 &   1.87 &  0.840E-03 \\
373385 &  15.10 &   3.164062 &   0.060718 &   0.161308 & 126.839681 & -81.528174 &   1.51 &  0.520E-03 \\
373392 &  15.91 &   3.156675 &   0.055445 &   0.156193 & 121.470180 & -81.141158 &   1.65 &  0.340E-03 \\
379091 &  15.93 &   3.164728 &   0.061350 &   0.161280 & 127.506180 & -81.618046 &   1.49 &  0.380E-03 \\
384154 &  16.04 &   3.160966 &   0.056247 &   0.159521 & 124.286864 & -81.141267 &   1.66 &  0.200E-04 \\
386237 &  15.90 &   3.164115 &   0.060201 &   0.158957 & 127.442750 & -81.671175 &   1.49 &  0.180E-03 \\
386253 &  16.15 &   3.164490 &   0.060548 &   0.163292 & 126.745194 & -81.369909 &   1.48 &  0.860E-03 \\
388363 &  15.58 &   3.164903 &   0.060665 &   0.163347 & 127.129947 & -81.394252 &   1.44 &  0.740E-03 \\
389221 &  16.49 &   3.165599 &   0.065974 &   0.162128 & 128.358144 & -82.038951 &   1.48 &  0.779E-03 \\
390971 &  16.14 &   3.159913 &   0.058079 &   0.155569 & 124.386518 & -81.560901 &   1.63 &  0.640E-03 \\
391728 &  16.06 &   3.165533 &   0.066185 &   0.162052 & 128.322252 & -82.063738 &   1.47 & -0.300E-03 \\
392556 &  16.31 &   3.163863 &   0.065132 &   0.155402 & 128.294777 & -82.417095 &   1.73 & -0.264E-02 \\
392580 &  16.05 &   3.162223 &   0.057426 &   0.160864 & 125.120510 & -81.189675 &   1.75 & -0.110E-02 \\
\hline
\end{longtable}
\end{center}
\twocolumn

\bsp

\label{lastpage}

\end{document}